\documentclass[twocolumn]{aastex61}

\received{April 7, 2017}
\accepted{May 30, 2017}
\submitjournal{ApJ}

\shorttitle{Mass transport from the envelope to the disk of V346 Nor}
\shortauthors{K\'osp\'al et al.}

\usepackage{color}

\begin{document}

\title{Mass transport from the envelope to the disk of V346 Nor: a
  case study for the luminosity problem in an FUor-type young eruptive
  star}

\author{\'A. K\'osp\'al} \affiliation{Konkoly Observatory, Research
  Centre for Astronomy and Earth Sciences, Hungarian Academy of
  Sciences, Konkoly-Thege Mikl\'os \'ut 15-17, 1121 Budapest, Hungary}
  \affiliation{Max Planck Institute for Astronomy, K\"onigstuhl 17,
  69117 Heidelberg, Germany} \email{kospal@konkoly.hu}
  
\author{P. \'Abrah\'am} \affiliation{Konkoly Observatory, Research
  Centre for Astronomy and Earth Sciences, Hungarian Academy of
  Sciences, Konkoly-Thege Mikl\'os \'ut 15-17, 1121 Budapest, Hungary}
  
\author{T. Csengeri} \affiliation{Max Planck Institute for
  Radioastronomy, Auf dem H\"ugel 69, 53121 Bonn, Germany}
 
\author{O. Feh\'er} \affiliation{Konkoly Observatory, Research Centre
  for Astronomy and Earth Sciences, Hungarian Academy of Sciences,
  Konkoly-Thege Mikl\'os \'ut 15-17, 1121 Budapest, Hungary}
  
\author{M. R. Hogerheijde} \affiliation{Leiden Observatory, Leiden
  University, P.O. Box 9513, 2300 RA Leiden, The Netherlands}
  
\author{Ch. Brinch} \affiliation{Niels Bohr International Academy, The
  Niels Bohr Institute, University of Copenhagen, Blegdamsvej 17, 2100
  Copenhagen {\O}, Denmark}
  
\author{M. M. Dunham} \affiliation{Department of Physics, State
  University of New York at Fredonia, Fredonia, NY 14063, USA}
\affiliation{Harvard-Smithsonian Center for Astrophysics, 60 Garden
  Street, MS 78, Cambridge, MA 02138, USA}
  
\author{E. I. Vorobyov} \affiliation{Department of Astrophysics, The
  University of Vienna, Vienna, A-1180, Austria} \affiliation{Research
  Institute of Physics, Southern Federal University, Stachki 194,
  Rostov-on-Don, 344090, Russia}

\author{D. M. Salter} \affiliation{Department of Astronomy and Laboratory
  for Millimeter-Wave Astronomy, University of Maryland, College Park,
  MD 20742, USA}
  
\author{Th. Henning} \affiliation{Max Planck Institute for Astronomy,
  K\"onigstuhl 17, 69117 Heidelberg, Germany}


\begin{abstract}
A long-standing open issue of the paradigm of low-mass star formation
is the luminosity problem: most protostars are less luminous than
theoretically predicted. One possible solution is that the accretion
process is episodic. FU~Ori-type stars (FUors) are thought to be the
visible examples for objects in the high accretion state. FUors are
often surrounded by massive envelopes, which replenish the disk
material and enable the disk to produce accretion outbursts. However,
we have insufficient information on the envelope dynamics in
FUors, about where and how mass transfer from the envelope to the disk
happens. Here we present ALMA observations of the FUor-type star
V346~Nor at 1.3\,mm continuum and in different CO rotational lines. We
mapped the density and velocity structure of its envelope and analyze
the results using channel maps, position-velocity diagrams, and
spectro-astrometric methods. We found that V346~Nor is surrounded by
gaseous material on 10\,000\,au scale in which a prominent outflow
cavity is carved. Within the central $\sim$700\,au, the circumstellar
matter forms a flattened pseudo-disk where material is infalling with
conserved angular momentum. Within $\sim$350\,au, the velocity profile
is more consistent with a disk in Keplerian rotation around a central
star of 0.1$\,M_{\odot}$. We determined an infall rate from the
envelope onto the disk of 6$\times$10$^{-6}\,M_{\odot}$\,yr$^{-1}$, a
factor of few higher than the quiescent accretion rate from the disk
onto the star, hinting for a mismatch between the infall and accretion
rates as the cause of the eruption.
\end{abstract}



\section{Introduction}
\label{sec:intro}

Sun-like stars form when dense cores in the interstellar matter
gravitationally collapse. Nascent stars are surrounded by
circumstellar disks, from which material is accreted onto the growing
star. Initially, the star+disk system is embedded in an envelope, the
remnant of the initial core, which feeds material to the disk. A
long-standing problem with this paradigm is the luminosity problem:
theoretical models for the collapse of cloud cores predict infall
rates on the order of 10$^{-6}\,M_{\odot}$\,yr$^{-1}$, which imply
luminosities typically 10-100 times higher than what is observed for
embedded protostars \citep[][and references therein]{dunham2014}. One
way to overcome this conundrum is to assume that the accretion rate is
not constant in time, but episodic: the protostar normally accretes at
a very low rate, and this quiescent accretion is occasionally
interspersed by brief episodes of highly enhanced accretion (Kenyon
1990). By combining radiative transfer with hydrodynamical
simulations, our group found that models predicting accretion rates
that both decline with time and feature short-term variability and
episodic bursts reproduce well the observed protostellar luminosity
function, offering a solution for the luminosity problem
\citep{dunham2012, dunham2013}.

FU~Orionis-type variables (FUors) are thought to be the visible
examples of episodic accretion. FUors exhibit 5-6\,mag optical
outbursts attributed to highly enhanced accretion \citep{hk96}. During
these outbursts, accretion rates from the circumstellar disk onto the
star are on the order of 10$^{-4}\,M_{\odot}$\,yr$^{-1}$, three orders
of magnitude higher than in quiescence. The exact physical mechanism
of FUor outbursts is debated. Explanations include viscous-thermal
instabilities in the disk \citep{bell1994}, a combination of
gravitational and magneto-rotational instability \citep{armitage2001},
or accretion of clumps in a gravitationally fragmenting disk
\citep{vorobyov2006, vorobyov2010, vorobyov2015}. Yet another type of
theory involves a close stellar or sub-stellar companion that perturbs
the disk and triggers the onset of the enhanced accretion
\citep{lodato2004, bonnell1992, nayakshin2012}.

Envelopes play a significant role in the outburst of FUors by
replenishing the disk material after each outburst and sustaining
gravitational instability, the key ingredient of several outburst
models \citep{zhu2009, vorobyov2010, vorobyov2015}. A fundamental
parameter is the mass infall rate from the envelope onto the disk,
which regulates all these processes and determines the periodicity of
disk fragmentation and, indirectly, the frequency of bursts caused by
clump infall \citep{vorobyov2013}. A general prediction of the
instability models is that the mass transport rate towards the inner
disk regions should exceed a critical value of a few times
10$^{-7}\,M_{\odot}$\,yr$^{-1}$, otherwise the material is accreted
steadily onto the star, and no eruptions are produced
\citep{bell1994}.  With a reasonable assumption that the mass infall
rate from the envelope onto the disk is of the same order as the mass
transport through the disk, this also implies the existence of a
critical envelope infall rate.  With ALMA, it is now possible to study
the infall process and characterize the FUor envelopes in details.

In this paper, we use new ALMA observations to analyze the structure
and properties of the circumstellar matter around V346~Nor, an
embedded FUor located in the Sandqvist 187 dark cloud, at a distance
of 700\,pc \citep{reipurth1981}. It went into eruption some time
between 1976 and 1980, reached a peak accretion rate of
10$^{-4}\,M_{\odot}$\,yr$^{-1}$ in 1992, produced an unexpected, rapid
fading in 2010-11, and after a minimum, brightened again
\citep{kraus2016, kospal2017a}. It is associated with the Herbig-Haro
object HH~57 \citep{reipurth1985}. \citet{sw2001} observed V346~Nor with
the JCMT/SCUBA in the submillimeter continuum, measuring a size of
22$''{\times}$8$''$ and a total (gas+dust) envelope mass of
0.5$\,M_{\odot}$. \citet{evans1994} detected very strong
$^{12}$CO(3--2) and $^{13}$CO(2--1) lines with the CSO, indicating an
outflow, and measured an envelope mass of about
1$\,M_{\odot}$. Recently, we also measured CO emission from V346~Nor,
using APEX for the $^{12}$CO(3--2), $^{12}$CO(4--3), and
$^{13}$CO(3--2) lines \citep{kospal2017b}. We obtained an envelope
mass of 0.3$\,M_{\odot}$ within a radius of 10\,000\,au (14$\farcs$3),
which is consistent with previous results for the extended envelope,
considering the uncertainties in mass determination. We also detected
the energetic outflow in the wings of the $^{12}$CO lines.

With ALMA, we performed continuum and CO line observations of V346~Nor
with better sensitivity, spatial, and velocity resolution than ever
before. In Section~\ref{sec:obs} we give details about our
observations and data reduction, in Section~\ref{sec:res} we present
our results, and in Section~\ref{sec:disc} we analyze the ALMA data
and discuss the observations in the context of what we know about the
infall process and envelope structure of Class\,0/I objects and other
FUors.


\section{Observations}
\label{sec:obs}

We observed V346~Nor using ALMA in Cycle 2 with the 12\,m array on
April 28, 2015, with the 7\,m (Morita) array on June 6, June 8, and
August 8, 2014, and with the Total Power antennas on June 14-15 and
July 16-18, 2015 (project 2013.1.00870.S, PI: \'A.~K\'osp\'al). The
12\,m array observations were executed with 37 antennas, using Titan
as flux calibrator, J1617-5848 and J1517-2422 as bandpass calibrators,
and J1623-4501 as phase calibrator. The 7\,m array observations were
done with 9-11 antennas using Mars and Titan as flux calibrators,
J1427-4206 as bandpass calibrator, and J1636-4102 as phase
calibrator. The Total Power observations were obtained with 3
antennas, and Uranus was used as a flux calibrator.

Tab.~\ref{tab:log} shows the log of observations. The interferometric
observations were single pointings, while the Total Power measurements
were on-the-fly maps of about 80$''{\times}$80$''$, with an OFF
position at $\Delta$RA = $-$19$\farcm$10 and $\Delta$DEC =
+44$\farcm$98, selected to be emission-free in CO at the expected
velocity of our science target.  The 12\,m array had projected
baseline lengths between 14\,m and 349\,m (11 -- 268\,k$\lambda$),
while the 7\,m array had projected baseline lengths between 8.3\,m and
48\,m (6 -- 37\,k$\lambda$). The observations were done in Band
6. Three spectral windows of 58.6\,MHz bandwidth and 30\,kHz
(40\,m\,s$^{-1}$) resolution were centered at the $^{12}$CO,
$^{13}$CO, and C$^{18}$O J=2--1 lines, while one spectral window of
1875\,MHz bandwidth was centered at 233\,GHz to measure the 1.3\,mm
continuum.

All data sets were calibrated using the standard ALMA reduction
software CASA (v4.5; \citealt{mcmullin2007}). The visibilities on the
baselines in common for both the 7\,m array and the 12\,m array
interferometric data sets were in good agreement, therefore they were
concatenated into a single measurement set. This data set was used to
create a continuum image by excluding channels contaminated by line
emission, and cleaning with Briggs weighting using a robustness
parameter of 0.5. For the line data, we first fitted a continuum to
the line-free channels and subtracted it in the $uv$ space, then
cleaned the images with Briggs weighting using a robustness parameter
of 0.5. Finally, Total Power data cubes were prepared with the same
spatial and spectral gridding as the interferometric data, and they
were combined together in the image space using CASA's ``feather''
algorithm. We emphasize that the combination of the three data sets
allows us to image the emission on all spatial scales without flux
loss.

\begin{table}
\begin{center}
\caption{Log of observations.\label{tab:log}}
\begin{tabular}{cccc}
\tableline\tableline
Image           & Frequency (GHz) & Beam size & Beam P.A. \\
\tableline
Continuum       & 226.70856 & 0$\farcs$90$\times$1$\farcs$11 & $-$85$\fdg$1 \\
$^{12}$CO J=2--1 & 230.53800 & 0$\farcs$91$\times$1$\farcs$11 & $-$86$\fdg$6 \\
$^{13}$CO J=2--1 & 220.39868 & 0$\farcs$95$\times$1$\farcs$16 & $-$85$\fdg$4 \\
C$^{18}$O J=2--1 & 219.56035 & 0$\farcs$97$\times$1$\farcs$19 & $-$86$\fdg$0 \\
\tableline
\end{tabular}
\end{center}
\end{table}


\section{Results and Analysis}
\label{sec:res}

\subsection{Continuum image}
\label{sec:cont}

\begin{figure*}
\includegraphics[angle=0,scale=.54]{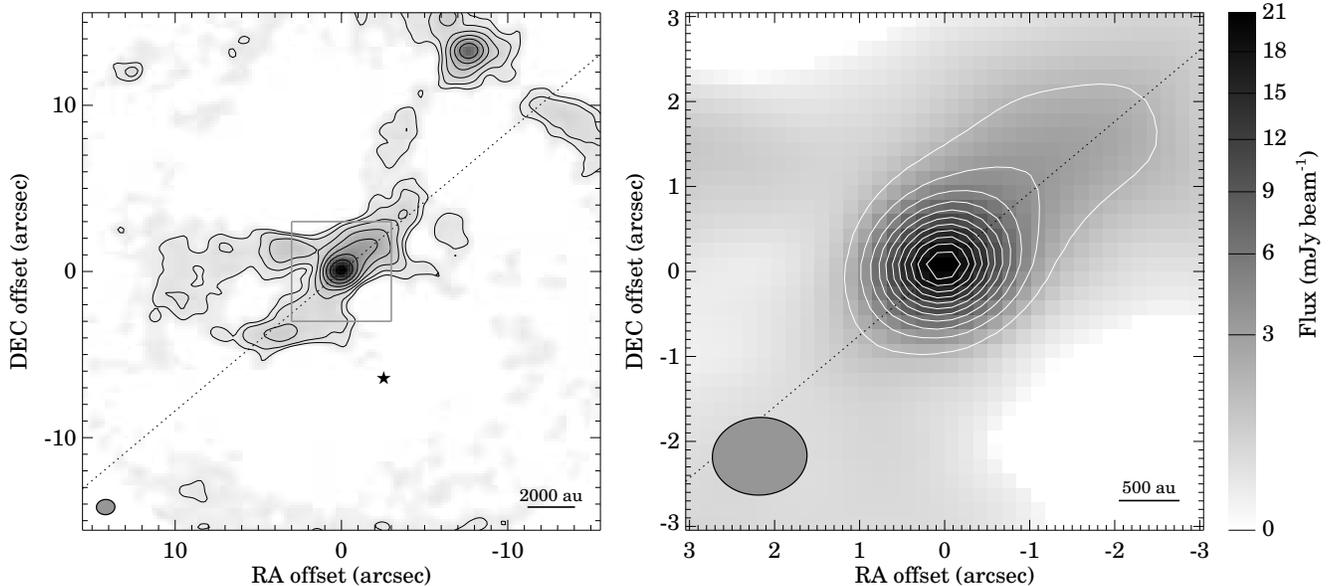}
\caption{ALMA 1.322\,mm continuum image of V346~Nor displayed with
  square root scaling. The contours in the left panel are at 2, 4, 8,
  16, 32, 64, 128, and 256\,$\sigma$, while in the right panel, the
  contours go from 25\,$\sigma$ to 275\,$\sigma$ with a 25\,$\sigma$
  spacing ($\sigma$ = 0.07\,mJy\,beam$^{-1}$). The asterisk marks the
  position of HH\,57. The dotted line displays the position angle of
  the disk seen in the CO data. \label{fig:cont}}
\end{figure*}

As Fig.~\ref{fig:cont} shows, V346 Nor is firmly detected at 1.322\,mm
continuum. The rms noise in the image is 0.07\,mJy\,beam$^{-1}$. The
peak flux is 19.7\,mJy\,beam$^{-1}$ (280\,$\sigma$ detection). The
peak is located at $\alpha_{2000}$=16$^{\rm h}$32$^{\rm m}$32$\fs$20,
$\delta_{2000}$=$-$44$^{\circ}$55$'$30$\farcs$69, which is only
0$\farcs$11 away from the phase center, and 0$\farcs$10 away from the
2MASS location of the central star in the near-infrared. Therefore,
the peak of the continuum emission coincides with the stellar
location. The central part of the continuum emission, displayed in
Fig.~\ref{fig:cont} right, is dominated by a fairly compact source.
The lower contours show an asymmetry: the continuum emission is
stronger in the northwestern direction, forming a finger-like, narrow
extension. This may be due to another, nearby source, some denser
clumps in the circumstellar envelope of V346~Nor, or a stream of
material falling toward the center. In an attempt to separate the
central peak from this extension, we fitted the brightness
distribution by two elliptical 2D Gaussians.

The fit revealed that the central source is marginally resolved, with
a deconvolved FWHM of 0$\farcs$46$\times$0$\farcs$60. It is elongated
along the northwest-southeast direction, with a position angle similar
to that of the rotating disk seen in the CO emission
(Sec.~\ref{sec:co}). At a distance of 700\,pc, its size corresponds to
a radius of 210\,au. The fitted flux of this structure is
27$\pm$3\,mJy, where the uncertainty already contains 10\% absolute
flux calibration uncertainty added in quadrature. Assuming
  optically thin emission, this can be converted to dust mass using
the following formula:
\begin{equation}
  m_{\rm dust} = \frac{f_{850}d^2}{\kappa_{850}B_{\nu}(T_{\rm dust})}
\end{equation}
We obtained a dust mass of 7$\times$10$^{-4}\,M_{\odot}$ using $d$ =
700\,pc for the distance of the source,
$\kappa_{850}$=0.17\,m$^2$\,kg$^{-1}$ for the dust opacity at
850$\,\mu$m \citep{zuckerman1993}, $T_{\rm dust}$ = 50\,K for the
temperature, and $\beta$=1 for the dust emissivity \citep{sw2001}. In
order to check our assumption about the optical depth $\tau$, we
calculated a $\tau$ map by comparing the observed emission with
blackbody radiation using 50\,K dust temperature. The resulting peak
$\tau$ was around 0.01, and even with $T_{\rm dust}$ = 20\,K, the
$\tau$ stays below 0.03. Therefore, our calculated dust mass do not
underestimate the true value. As we will show in
Sec.~\ref{sec:specast}, the Keplerian disk around the star
approximately fills the central beam, thus, no significant beam
dilution is expected, further supporting our mass estimate.  According
to our fit, the nearby extended source to the northwest is at a
distance of 1$\farcs$9 from V346~Nor and is elongated in the same
direction with a deconvolved FWHM of 1$\farcs$9. Its peak brightness
is about 7 times fainter than that of V346~Nor. Therefore, V346~Nor
dominates the continuum emission within about $\sim$1$''$ radius.

There is a significant amount of even fainter (but still detectable at
a high signal-to-noise ratio), extended emission around the
target. The lowest contour levels in Fig.~\ref{fig:cont} left (2, 4,
and 8\,$\sigma$) hint at four protrusions/tongues, which may indicate
the walls of an outflow cavity. We found that an aperture radius of
9$''$ (6300\,au) includes all these features. The total millimeter
flux within this radius is 42\,mJy. Subtracting from this the flux of
the central source, 27\,mJy, we computed the mass of the extended part
assuming 20\,K for the temperature and $\beta$=2 for the dust
emissivity \citep{sw2001}. The resulting dust mass for the extended
component is 1.6$\times$10$^{-3}\,M_{\odot}$. Adding to that the mass
of the central source, we obtained 2.3$\times$10$^{-3}\,M_{\odot}$ as
the total dust mass within 6300\,au.

There is an additional continuum source visible in the ALMA image,
located 15$\farcs$2 to the northwest of V346~Nor, at
$\alpha_{2000}$=16$^{\rm h}$32$^{\rm m}$31$\fs$48,
$\delta_{2000}$=$-$44$^{\circ}$55$'$17$\farcs$50, with a peak flux of
19.9\,mJy and a total flux of 27$\pm$7\,mJy. This source is unknown in
the literature, although it is clearly visible in Herschel 70$\,\mu$m
and 100$\,\mu$m images. It is more difficult to discern it at longer
wavelengths due to the degrading spatial resolution of Herschel. The
Herschel/PACS Point Source Catalogue \citep{marton2017} gives a flux of
4.1$\pm$0.4\,Jy and 7.4$\pm$0.4\,Jy at 70 and 100$\,\mu$m,
respectively. The source is invisible at optical and near-infrared
wavelengths, thus, it is probably a deeply embedded protostar.

\subsection{CO spectra}
\label{sec:co}

\begin{figure}
  \includegraphics[angle=90,scale=.57]{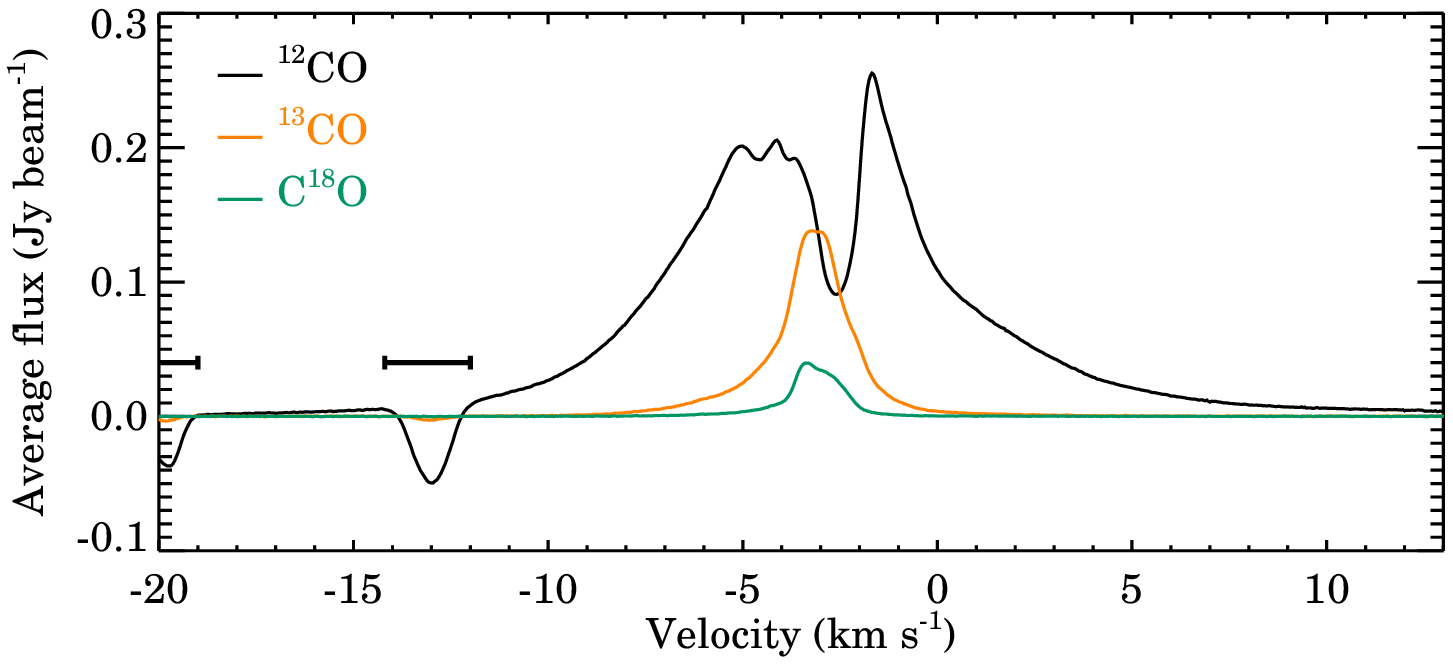}
  \includegraphics[angle=90,scale=.57]{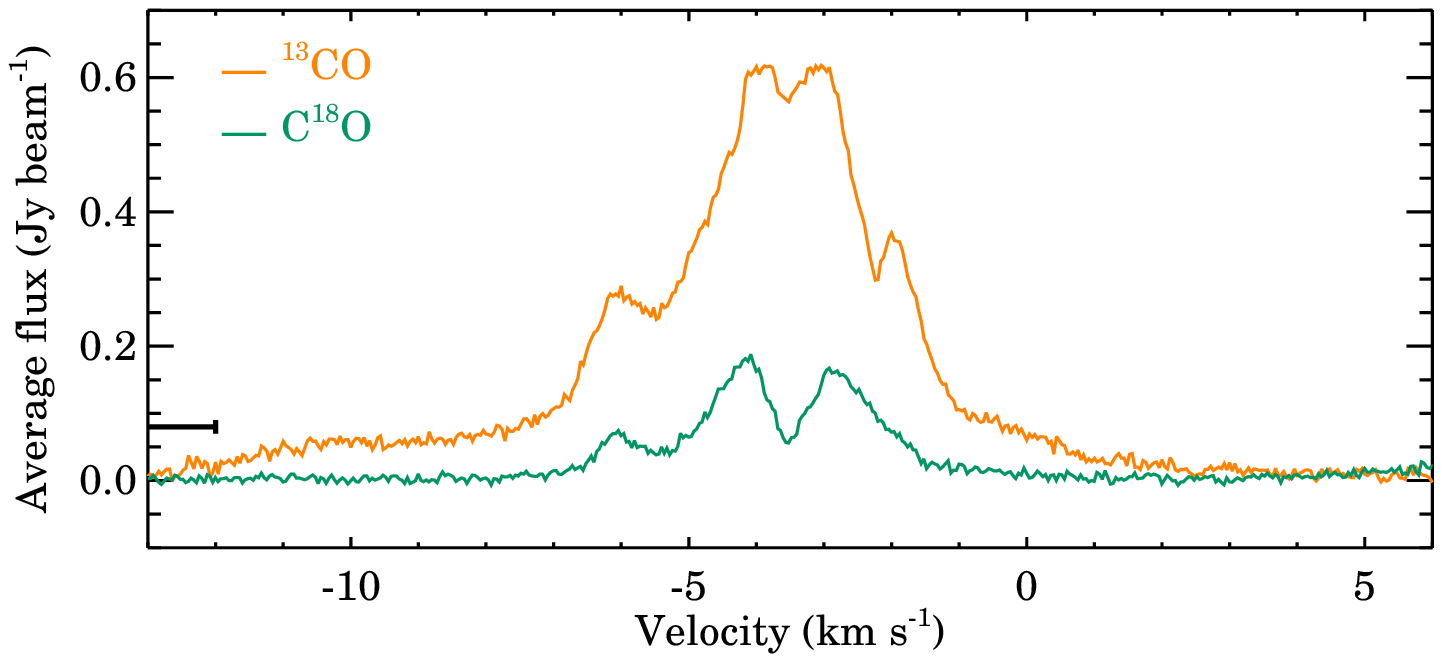}
  \caption{Spectra of the different CO isotopologues for the total
    field of view ($\approx$\,14\,000 au radius, top) and for the
    central beam ($\approx$\,700\,au, bottom). Black horizontal lines
    mark velocity ranges affected by emission in the OFF position in
    the total power maps. These artifacts are not present in the
    interferometric-only data. Vertical dotted lines mark the velocity
    ranges for low-velocity (LV), medium velocity (MV) and
    high-velocity (HV) gas, for which the integrated redshifted and
    blueshifted emission maps are displayed in
    Fig.~\ref{fig:CO_zoomin}.
  \label{fig:spectra}}
\end{figure}

Fig.~\ref{fig:spectra} shows the line profile of the three CO
isotopologues for an area with a radius of 14\,000\,au (20$''$), and
the two rarer isotopologues for the central beam (the $^{12}$CO
spectrum for the central beam suffers heavily from self-absorption,
and will not be discussed here). Negative features, unrelated to the
source, can be seen between $-$12 and $-$14.5\,km\,s$^{-1}$ and
between $-$19 and $-$21.5\,km\,s$^{-1}$. These are artifacts due to
emission in the OFF position in the total power maps, and they are not
present in the interferometric-only data. These velocity ranges are
omitted from our analyses. The spectra of the larger area show
profiles similar to what is typical for infalling envelopes around
highly embedded objects \citep{evans1999}. The optically thinner
$^{13}$CO and C$^{18}$O lines peak around the systemic velocity (see
below), while the optically thicker $^{12}$CO has strong
self-absorption at a velocity slightly redshifted from the systemic
velocity. This kind of line profile is explained in the literature as
an envelope with a radial velocity gradient: the outer part of the
envelope is mostly static, causing the absorption feature, while the
inner part is infalling toward the central star \citep[][and
  references therein]{shu1977,evans2015}.

The $^{13}$CO and C$^{18}$O spectra of the central beam
(Fig.~\ref{fig:spectra}, bottom) show two prominent peaks close to the
line center, symmetrically around a dip at $-$3.55\,km\,s$^{-1}$,
which we will adopt as the systemic velocity. The double-peaked line
profiles may indicate rotation in a compact source. There is a clear
indication of line wings, especially well visible in the stronger
$^{13}$CO line, up to $\pm$9\,km/s from the systemic velocity. This
may indicate outflows launched very close to the central star, or
material rotating at relatively high velocities. Based on the
appearance of these line profiles, we defined three velocity ranges
symmetrically around the systemic velocity: low velocity (LV) within
1.5\,km\,s$^{-1}$, medium velocity (MV) between
1.5...3.5\,km\,s$^{-1}$, and high velocity (HV) between
3.5...8.5\,km\,s$^{-1}$. These velocity ranges are marked in
Fig.~\ref{fig:spectra} as well. We integrated the CO emission in these
velocity ranges, and plotted the redshifted and blue-shifted emission
as contours in Fig.~\ref{fig:CO_zoomin}. In the Appendix, we also show
selected channel maps, in the case of $^{12}$CO for channels within
$\pm$16\,km\,s$^{-1}$ of the systemic velocity, while in the case of
the fainter $^{13}$CO and C$^{18}$O within $\pm$8\,km\,s$^{-1}$.

\subsection{CO maps}

\begin{figure*}
\includegraphics[angle=0,scale=.39]{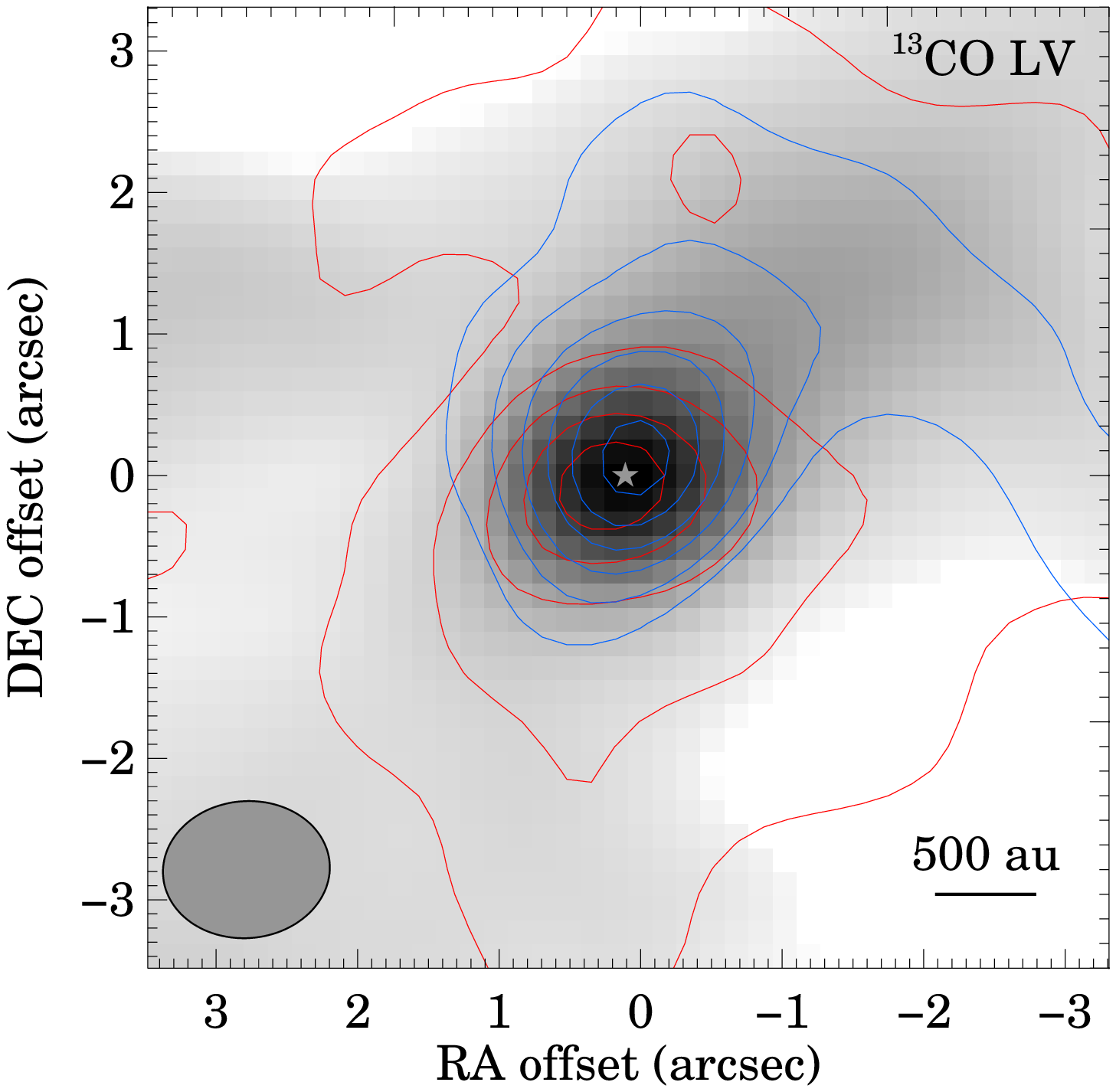}
\includegraphics[angle=0,scale=.39]{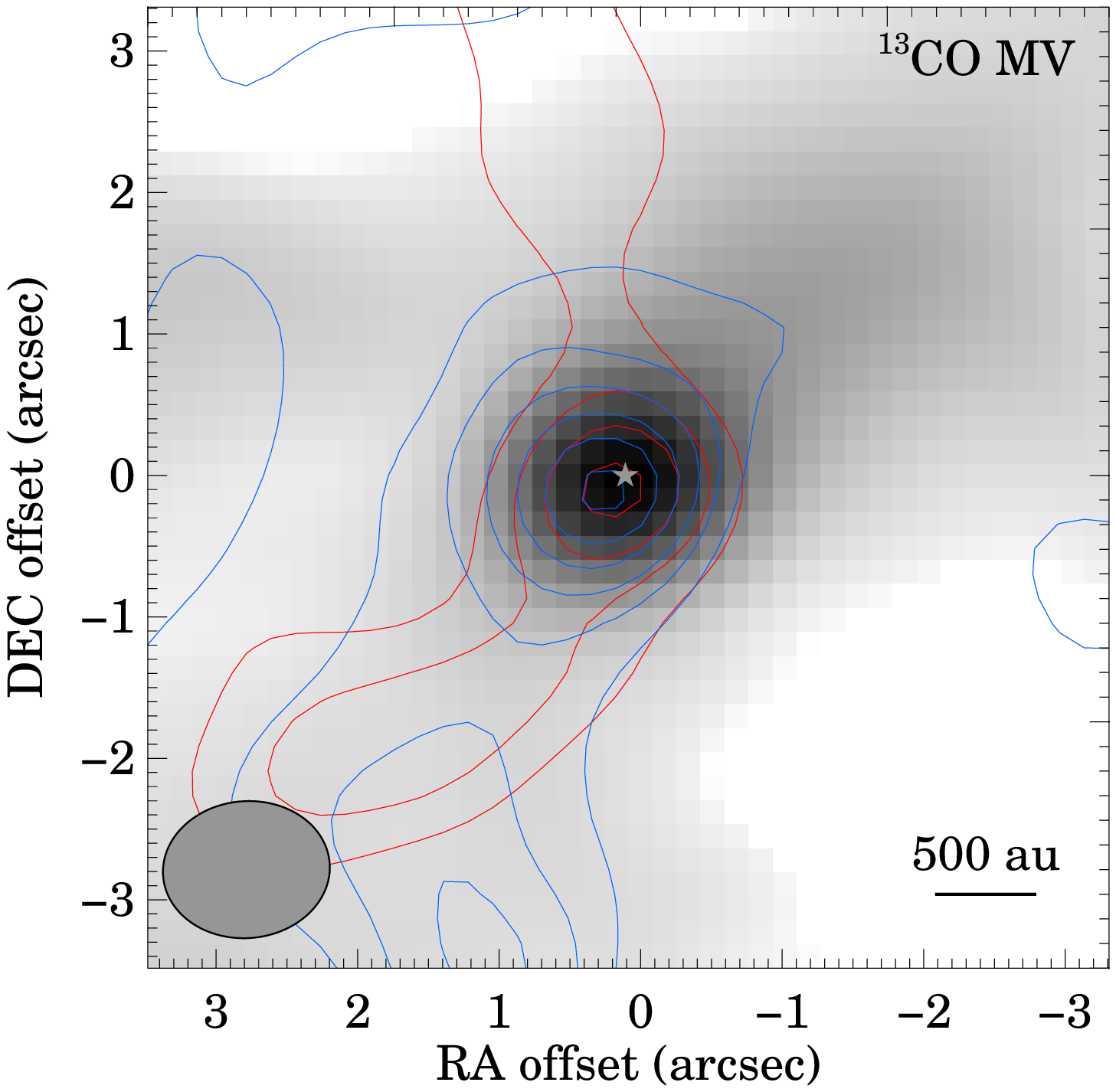}
\includegraphics[angle=0,scale=.39]{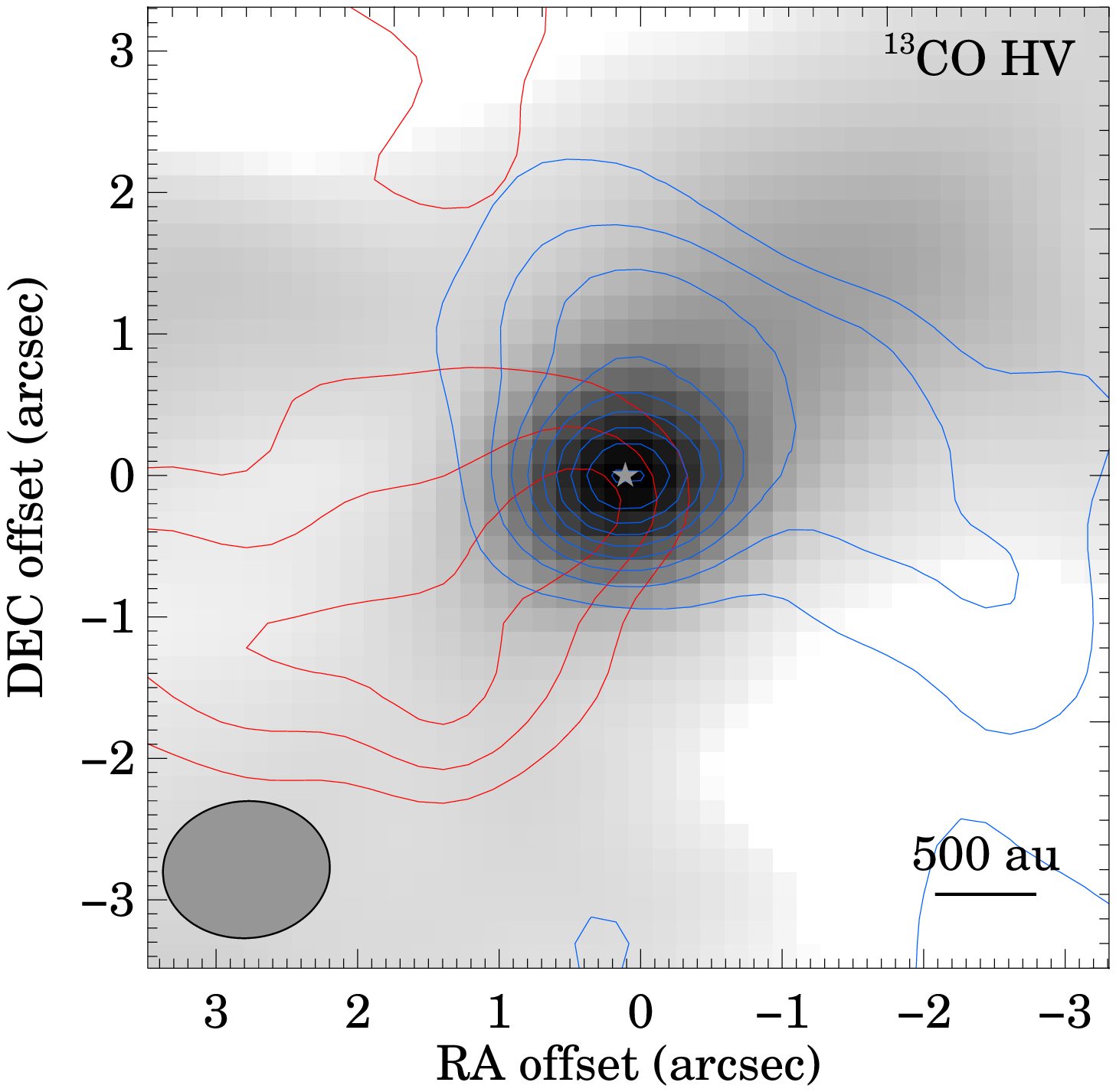}\\
\includegraphics[angle=0,scale=.39]{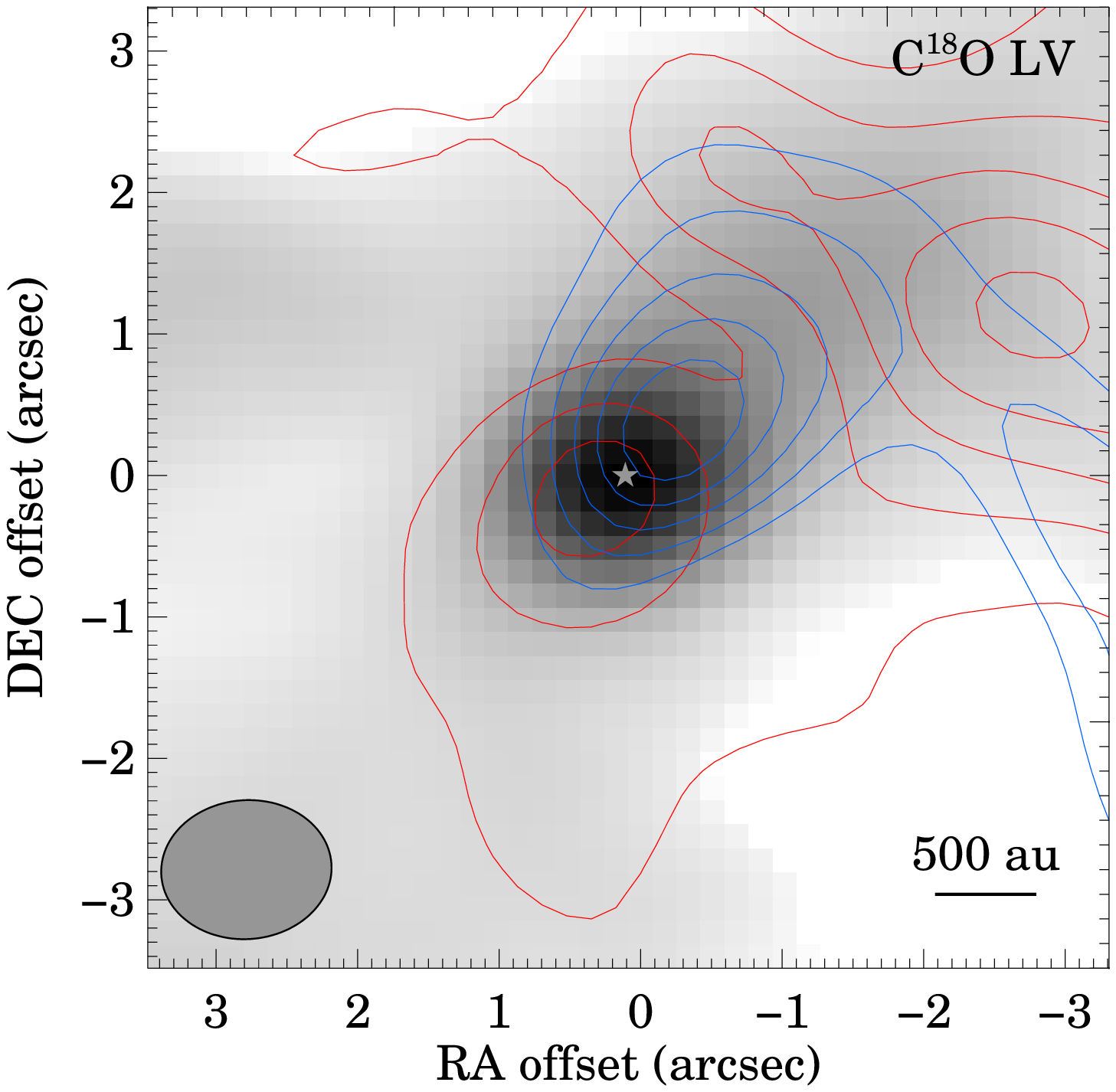}
\includegraphics[angle=0,scale=.39]{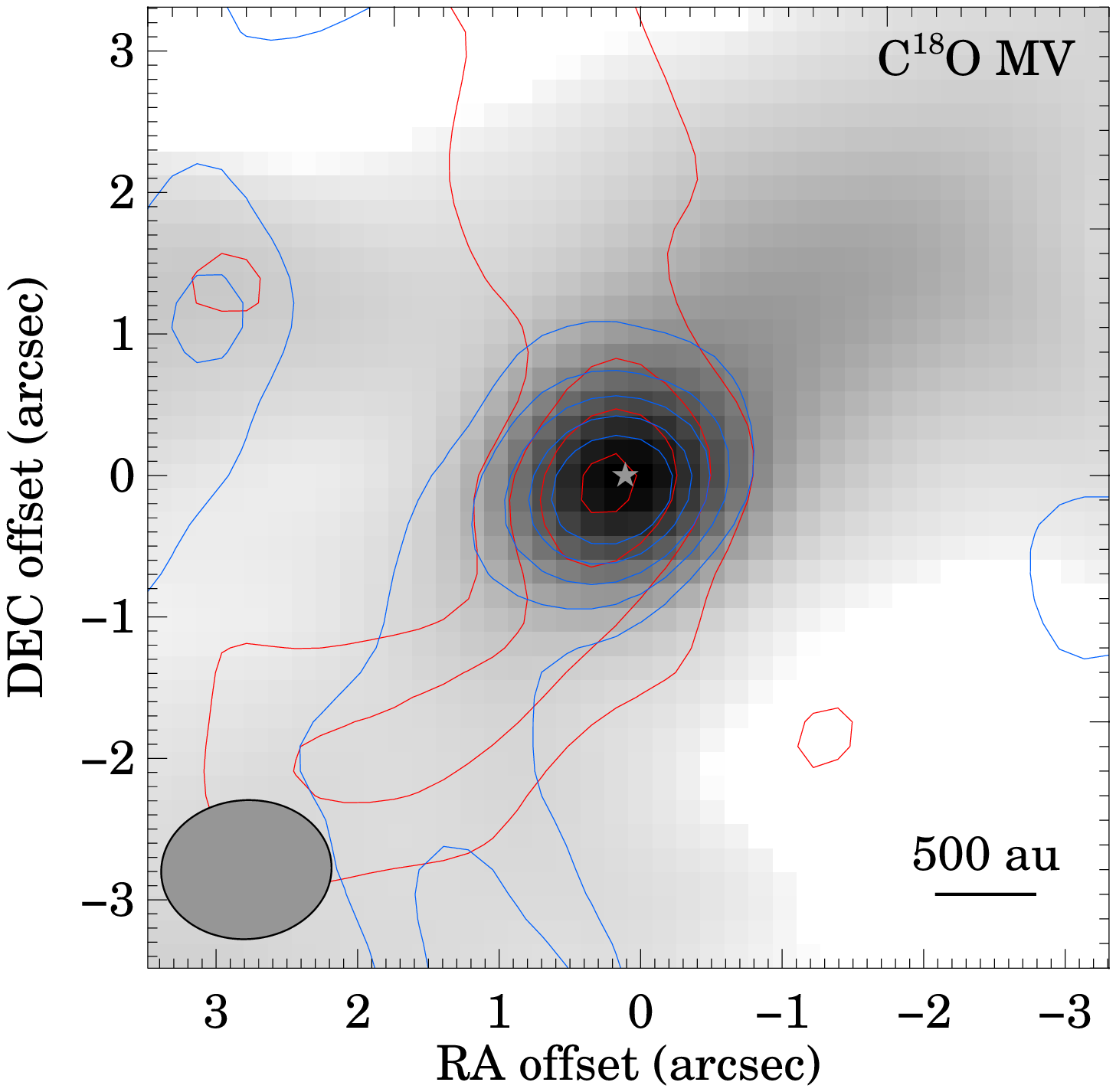}
\includegraphics[angle=0,scale=.39]{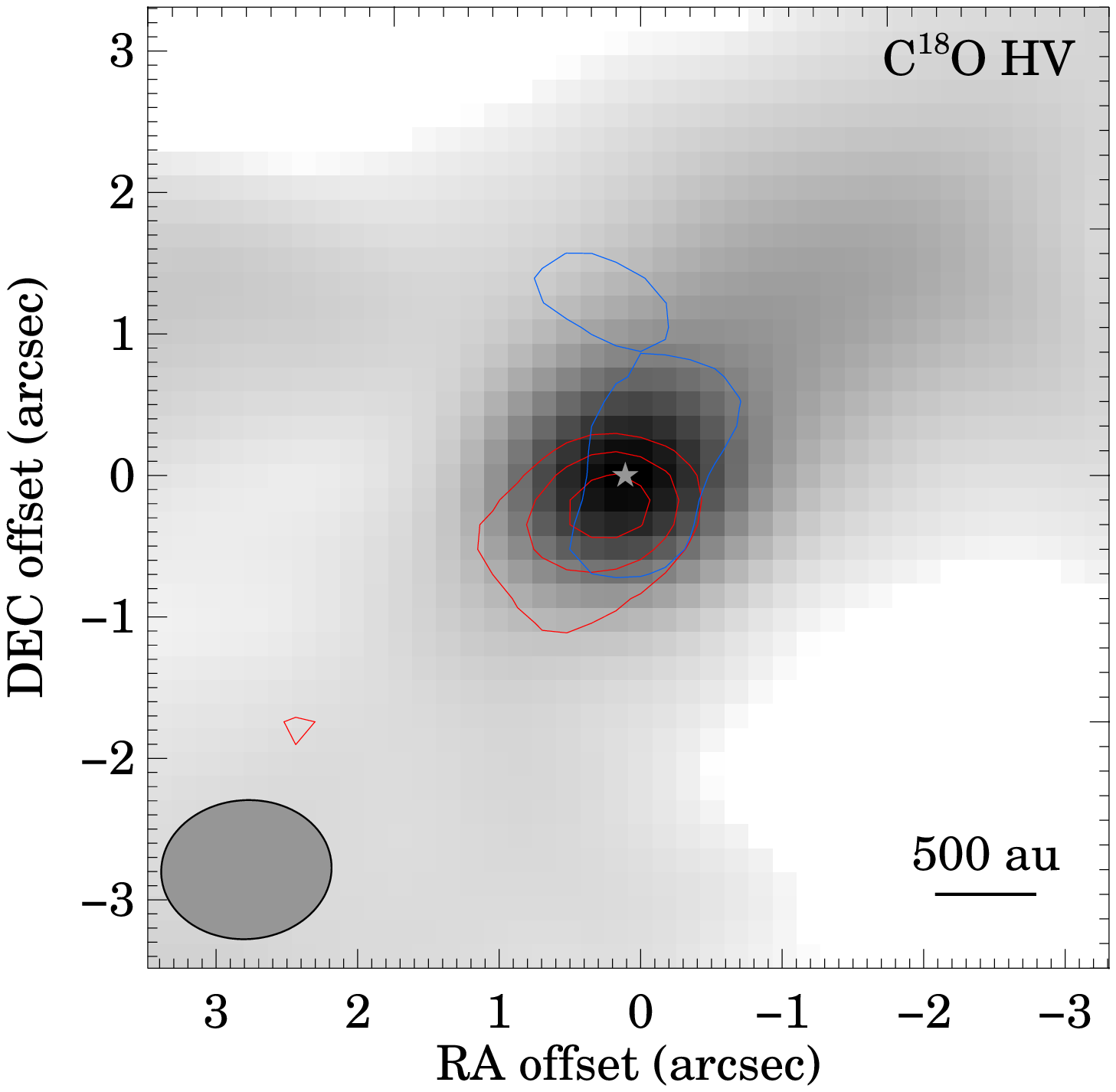}
\caption{Redshifted and blueshifted $^{13}$CO and C$^{18}$O emission
  of V346~Nor (red and blue contours) plotted over the 1.322\,mm
  continuum emission (grayscale). The gray asterisk marks the
  continuum peak. The beam is indicated in the lower left panel. The
  CO emission was integrated in the LV, MV, and HV velocity ranges
  marked in Fig~\ref{fig:spectra}. For $^{13}$CO LV the contours start
  at 60$\,\sigma$ with 20$\,\sigma$ spacing, for C$^{18}$O LV the
  contours start at 40$\,\sigma$ with 20$\,\sigma$ spacing, for
  $^{13}$CO MV the contours start at 30$\,\sigma$ with 10$\,\sigma$
  spacing, for C$^{18}$O MV the contours start at 10$\,\sigma$ with a
  5$\,\sigma$ spacing, for $^{13}$CO HV the contours start at
  6$\,\sigma$ with 4$\,\sigma$ spacing, and for C$^{18}$O HV the
  contours start at 4$\,\sigma$ with 2$\,\sigma$
  spacing. $\sigma$=5\,mJy\,beam$^{-1}$\,km\,s$^{-1}$ for $^{13}$CO
  LV, $\sigma$=2\,mJy\,beam$^{-1}$\,km\,s$^{-1}$ for C$^{18}$O LV,
  $\sigma$=7\,mJy\,beam$^{-1}$\,km\,s$^{-1}$ for $^{13}$CO MV,
  $\sigma$=3\,mJy\,beam$^{-1}$\,km\,s$^{-1}$ for C$^{18}$O MV,
  $\sigma$=10\,mJy\,beam$^{-1}$\,km\,s$^{-1}$ for $^{13}$CO HV, and
  $\sigma$=4\,mJy\,beam$^{-1}$\,km\,s$^{-1}$ for C$^{18}$O HV.
  \label{fig:CO_zoomin}}
\end{figure*}

\begin{figure}
\includegraphics[angle=0,scale=.555]{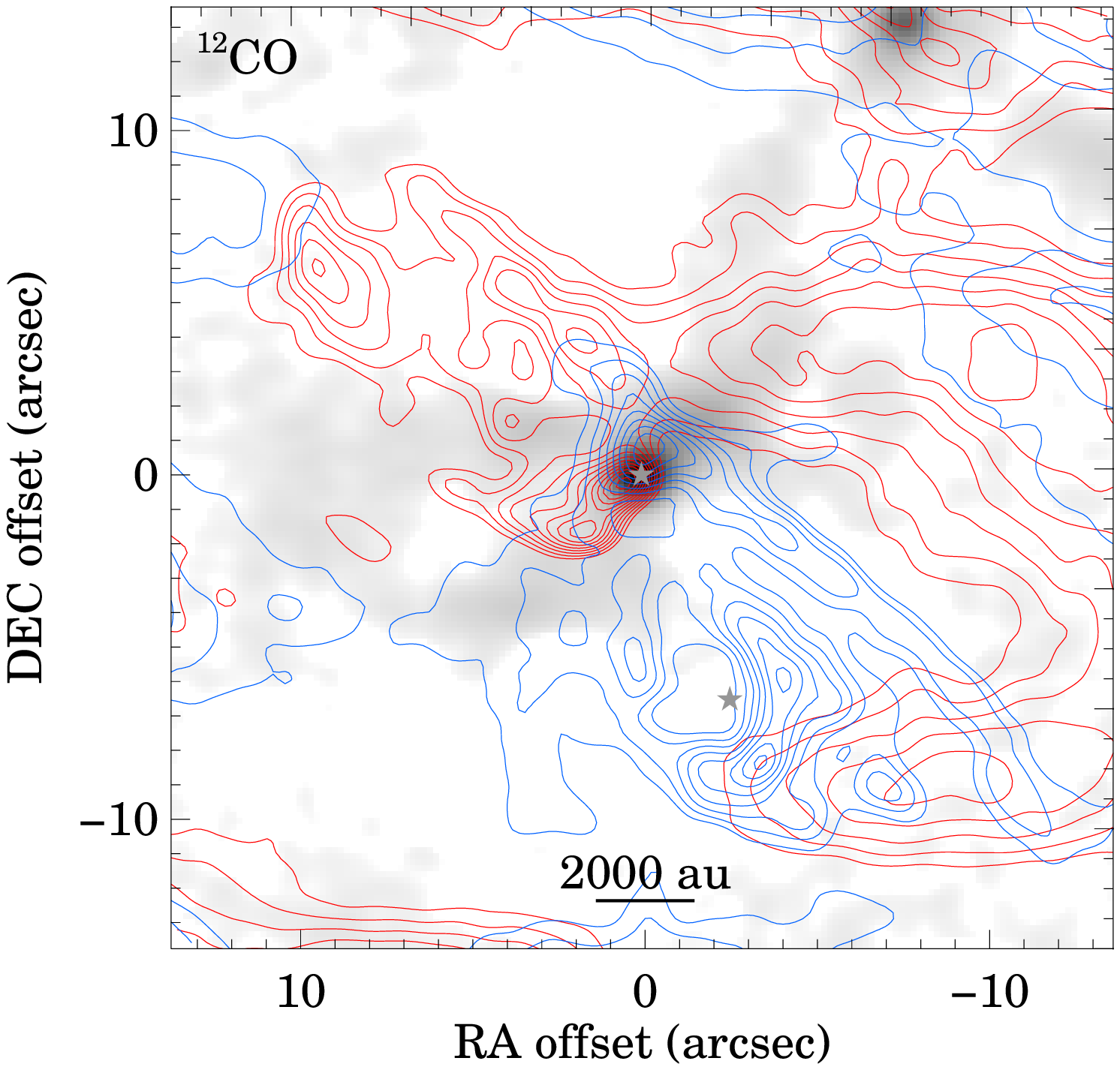}
\caption{Redshifted (from $-$1.5 to 15.9\,km\,s$^{-1}$) and
  blueshifted ($-$23 to $-$5.6\,km\,s$^{-1}$) $^{12}$CO emission of
  V346~Nor (red and blue contours), plotted over the 1.322\,mm
  continuum emission (grayscale). Velocity channels affected by
  absorption from the OFF position (Fig.~\ref{fig:spectra}) were
  discarded. Contours start at 250\,$\sigma$ with 250\,$\sigma$
  spacing, where $\sigma$ = 2\,mJy\,beam$^{-1}$\,km\,s$^{-1}$.
  \label{fig:CO_zoomout}}
\end{figure}

In Fig.~\ref{fig:CO_zoomin} the LV emission is quite extended, but
shows a centrally concentrated peak. The emission farther from the
image center is probably coming from the gaseous envelope extending
out to at least 10$''$ from the central star, clearly seen in the
channel maps in Figs.~\ref{fig:channelmaps_12co},
\ref{fig:channelmaps_13co}, and \ref{fig:channelmaps_c18o}. The
emission towards the center in Fig.~\ref{fig:CO_zoomin} shows a slight
offset between the blueshifted and redshifted centroids (redshifted is
offset toward the southeast, blueshifted is offset to the northwest
compared to the stellar position). This is clearly visible in
$^{13}$CO, but even more pronounced in C$^{18}$O. This suggests a
small rotating structure no larger than about 2$''$ in radius.

The MV emission is more compact. The maps show a structure where
emission is mostly concentrated along narrow clumpy filaments with
multiple bright peaks. There is a peak centered on the stellar
position, and we see no offset between the blueshifted and redshifted
centroids in Fig.~\ref{fig:CO_zoomin}. This result may reflect that in
case of Keplerian rotation, the higher velocity material is closer to
the central star. Allowing for the difference in signal-to-noise
ratio, the $^{13}$CO and C$^{18}$O maps are very similar an
appearance. The redshifted emission shows a parabola opening towards
the northeast with a relatively wide opening angle of about
80$^{\circ}$, while the blueshifted emission looks like a somewhat
narrower ellipse extending towards the southwest with an opening angle
of about 40$^{\circ}$. These structures are best visible at the
appropriate velocities in the $^{13}$CO channel maps in
Fig.~\ref{fig:channelmaps_13co}. This geometry is very reminiscent of
an outflow cavity, where emission is coming from the swept-up material
in the cavity walls. HH~57 seems to be situated along the axis of the
southwestern CO-emitting ellipse, close to its farther edge.

The HV emission (the first few and last few channels in
Figs.~\ref{fig:channelmaps_13co} and \ref{fig:channelmaps_c18o}) is
very similar in structure to the MV emission, so part of it is
possibly tracing the outflow cavity walls. The strong blueshifted
emission to the southwest surrounds the position of the Herbig-Haro
object HH~57, therefore, the strong CO emission may indicate the place
where the jet hits the ambient interstellar material. At these
velocities, the central part (Fig.~\ref{fig:CO_zoomin}) again shows an
offset between the redshifted and blueshifted emission. This is
unexpected for the rotating structure mentioned before. We propose
that instead, we may see a large-scale rotation at the base of the
parabolic/elliptic outflow cavities. Remarkable, the direction of
rotation is the same for both sides of the bipolar outflow and for the
compact central rotating object as well.

Fig.~\ref{fig:channelmaps_12co} in the Appendix reveals the structure
of the $^{12}$CO emission at even higher velocities, as far as
$\pm$15\,km\,s$^{-1}$ from the systemic velocity. In the blueshifted
part, a curious new feature appears as a very straight line pointing
from the star toward the southwest. In addition, the southwest
elliptical structure seen in the MV and HV maps in $^{13}$CO and
C$^{18}$O is also discernible in $^{12}$CO. On the redshifted side, a
new ellipse appears to the northeast. This seems to be different from
the parabolic structure seen in $^{13}$CO and C$^{18}$O, as it appears
at higher velocities, and has a narrower opening angle of about
40$^{\circ}$, although its position angle is similar. The red and
blueshifted emission displayed in Fig.~\ref{fig:CO_zoomout},
integrated for a very wide velocity range to account for the broad
$^{12}$CO J=2--1 line wings, also show these structures, but with
higher spatial resolution than any previous data, such as the
single-dish $^{12}$CO J=3--2 maps published in \citet{evans1994} and
\citet{kospal2017b}.

\subsection{Spectro-astrometry}
\label{sec:specast}

\begin{figure*}
\includegraphics[angle=90,scale=.5]{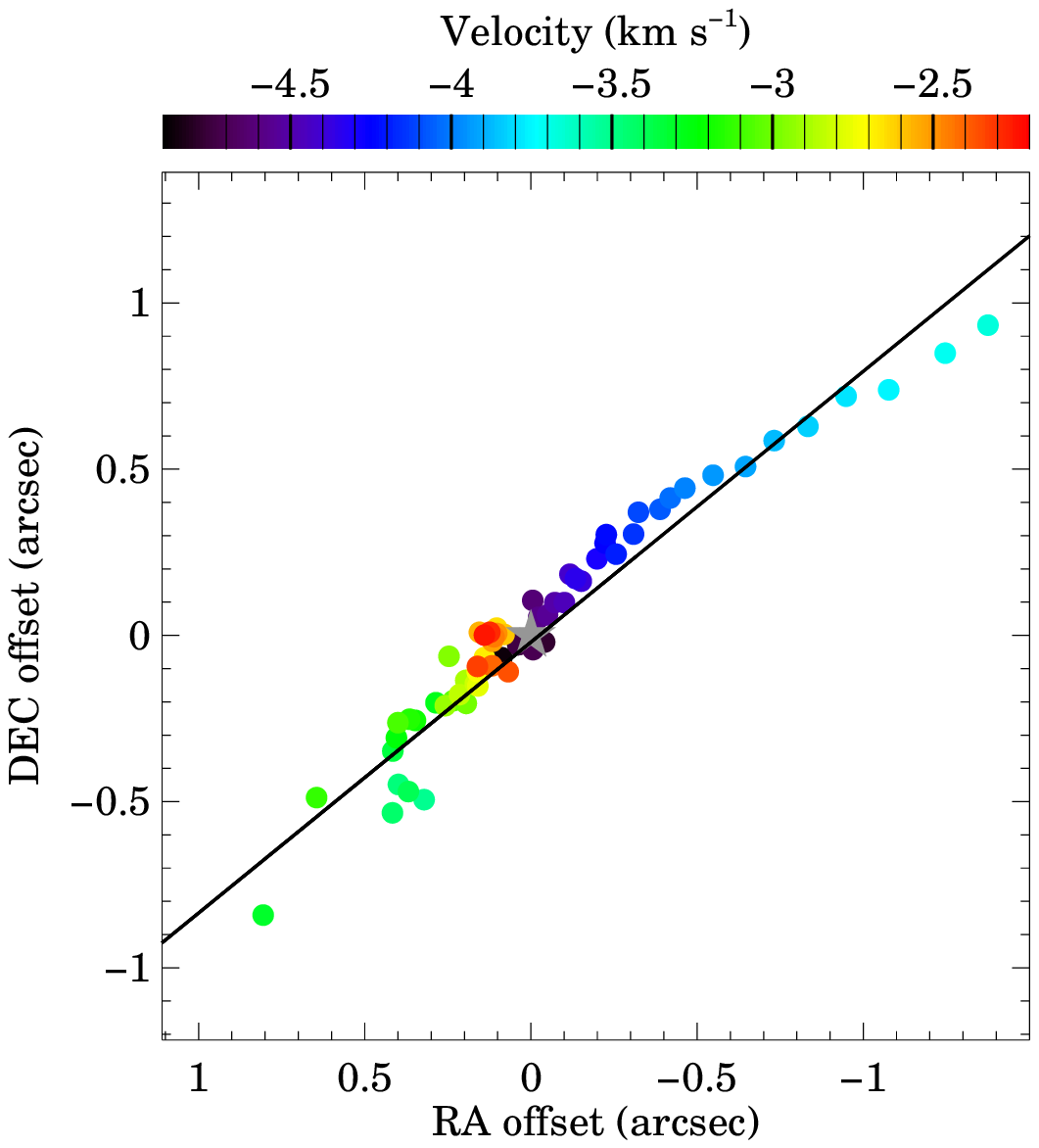}
\includegraphics[angle=90,scale=.5]{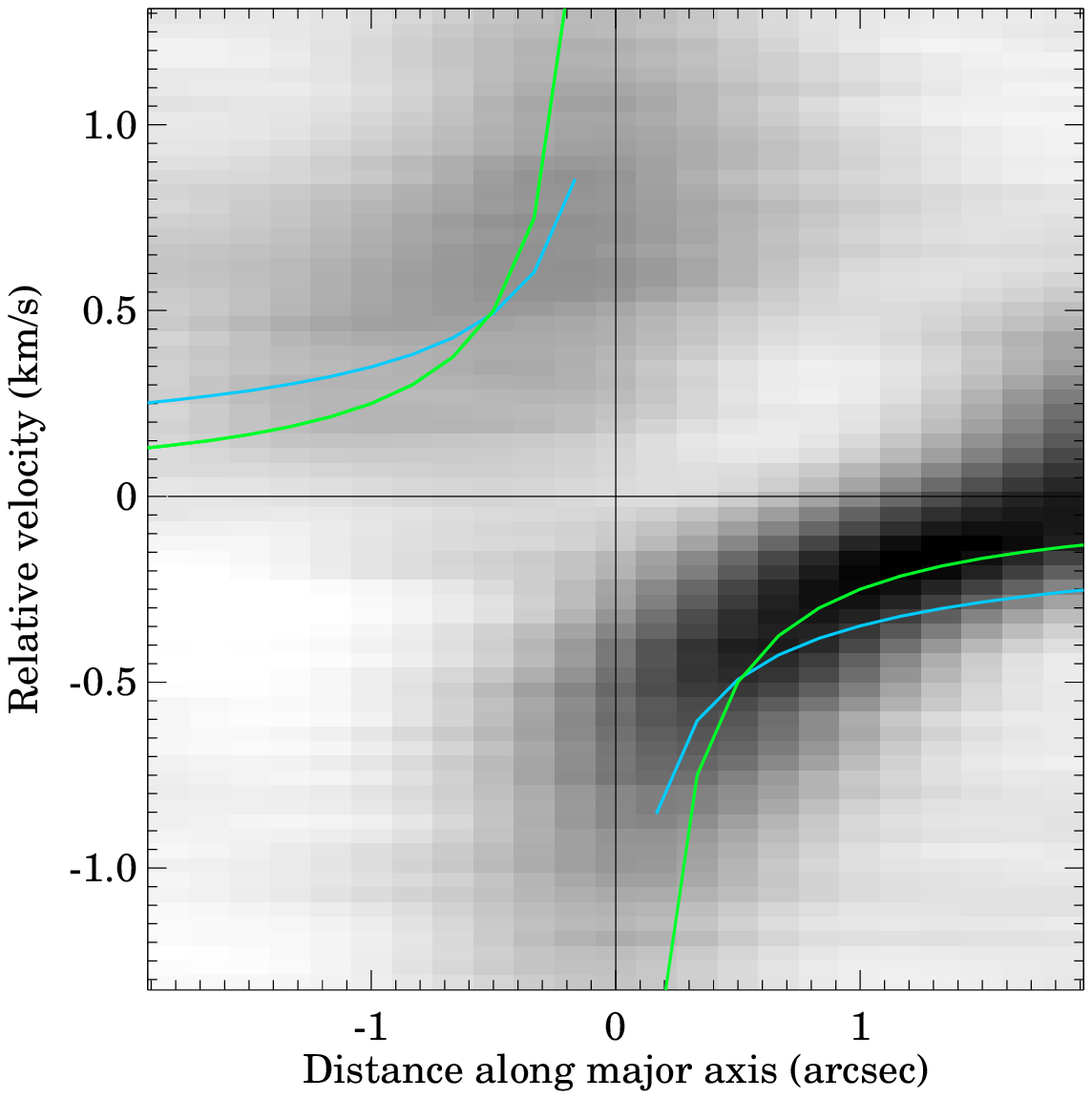}
\includegraphics[angle=90,scale=.5]{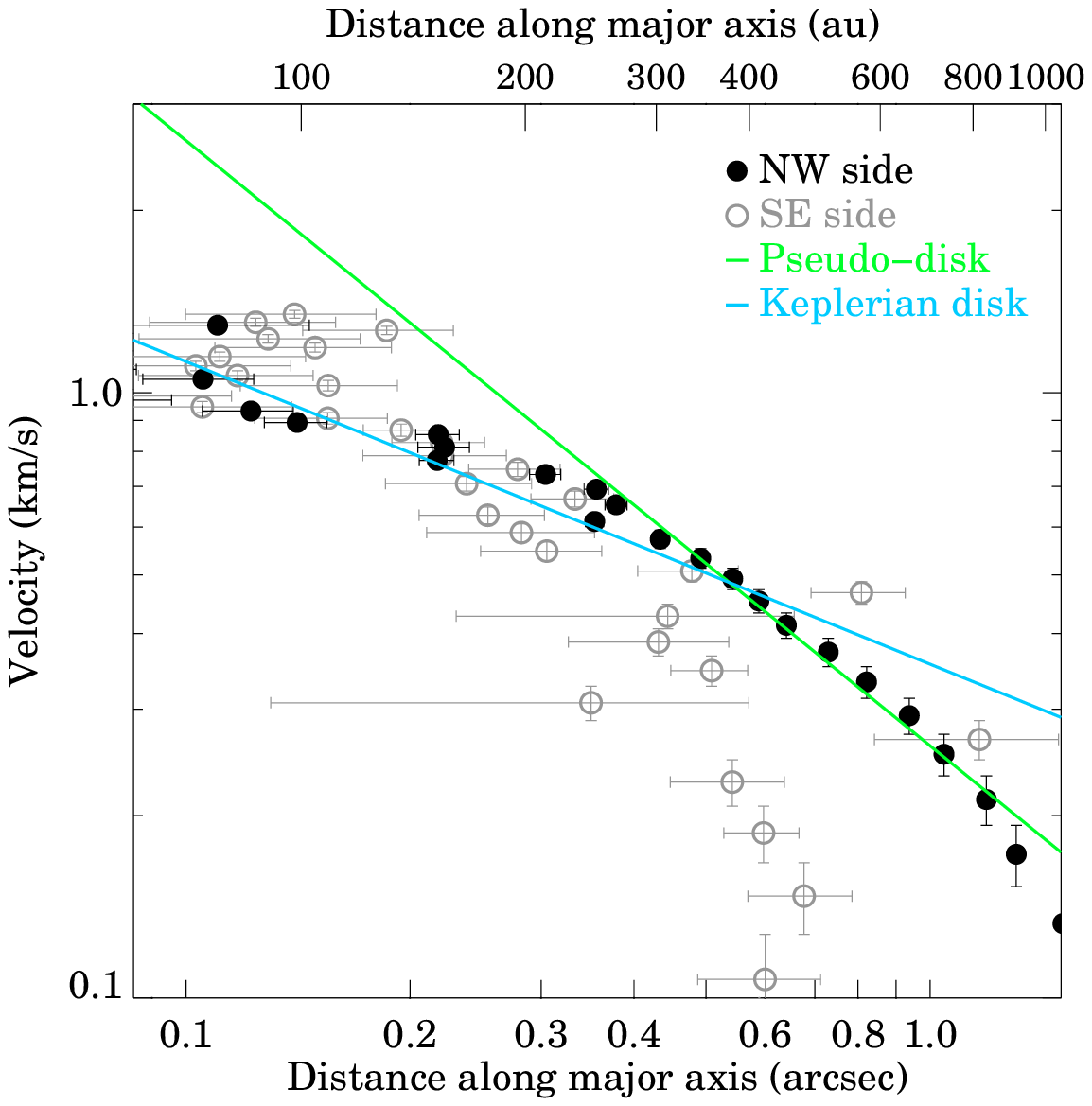}
\caption{Left: spectro-astrometric signal (centroid position for the
  different channels) for the C$^{18}$O data cube. The black line is a
  fit to the points, while the gray asterisk marks the location of the
  continuum peak. Middle: C$^{18}$O position-velocity diagram,
  relative to the systemic velocity of $-$3.55\,km\,s$^{-1}$. Right:
  velocity shift as a function of distance from the star for the
  C$^{18}$O data cube. The green curve is the velocity profile of a
  pseudo-disk ($\sim{}r^{-1}$), while the blue curve is a Keplerian
  profile ($\sim{}r^{-0.5}$).
  \label{fig:pvd}}
\end{figure*}

The rotation seen close to the central star at low velocities gives us
the opportunity to further study this compact rotating structure and
quantify its properties. We performed spectro-astrometry in our
C$^{18}$O data cube by calculating the centroid of the emission for
each velocity channel. The centroid positions lined up very precisely
along a position angle of 129$\pm$3$^{\circ}$ east of north
(Fig.~\ref{fig:pvd}). This outlines the direction of the major axis
perpendicular to the rotation axis. Next, we plotted a
position-velocity diagram along this major axis. Within about 1$''$,
this diagram looks as it is expected for a rotating disk. The
northwestern part is brighter, just like in the continuum map.

In Fig.~\ref{fig:pvd} we also plotted the velocities as a function of
the distance along the major axis as determined from the centroid
position for each channel map. We overplotted two possible
interpretations: a Keplerian disk with a radial profile of $r^{-0.5}$
(blue curve), and a pseudo-disk corresponding to infall with angular
momentum conservation, with a radial profile of $r^{-1}$ (green
curve). It seems that the inner part of the disk (within about
0$\farcs$5) is Keplerian, while the outer part is sub-Keplerian and is
more consistent with a pseudo-disk. This is mostly based on the
brighter northwestern part of the disk, because the southeastern part
is fainter and noisier, although a similar trend is apparent in the
southeastern part as well. The deviation from the $r^{-1}$ curve at
distances $>$1$''$ in the northwest side might be caused by
contamination from the nearby source or stream of material seen in the
continuum map and described in Sect.~\ref{sec:cont}. Graphs showing
similar transition from Keplerian to pseudo-disk with increasing
distance were found, e.g., in Class~I sources in Taurus by
\citet{harsono2014}.

The stellar mass of the best-fitting Keplerian curve is
0.1\,M$_{\odot}$, corresponding to a rather low mass central
star. This value is independent of the inclination, because it was
measured along the major axis. We estimate an uncertainty of the
stellar mass of about 20\%, including the uncertainty coming from the
distance, 700$\pm$100\,pc. From the pseudo-disk fit, we can derive the
angular momentum of the disk, which is 183\,au\,km\,s$^{-1}$
(cf.~168\,au\,km\,s$^{-1}$ for L1551\,IRS\,5,
\citealt{momose1998}). We estimated the mass of the disk by
integrating the $^{13}$CO line flux in a circular aperture within
0$\farcs$5 and within 1$\farcs$0, and obtained 0.01 and
0.03$\,M_{\odot}$, respectively. We assumed that the emission is
optically thin and assuming 50\,K average gas temperature, and taking
an isotopic ratio of 69 between $^{12}$CO and $^{13}$CO
\citep{wilson1999} and a canonical CO-to-H$_2$ abundance ratio of
10$^{-4}$. Mass estimates from C$^{18}$O give consistent results, a
posteriori validating our assumption about the low optical depth both
in the $^{13}$CO and C$^{18}$O lines. The value obtained for the
smaller area can be compared with the mass estimate from the continuum
for the central source. Assuming a dust-to-gas mass ratio of 100, the
total mass from the continuum measurement is 0.07$\,M_{\odot}$, while
it is only 0.01$\,M_{\odot}$, if calculated from the CO. This hints
for possible CO depletion (due to, e.g., freeze-out or conversion into
other species) in the center of the system. Similar results were
obtained also by \citet{jorgensen2002} for low-mass protostellar
envelopes and by \citet{murillo2013b} in the VLA\,1623 system.


\section{Discussion}
\label{sec:disc}

\subsection{Comparison to Class~0/I protostars}

Our observations revealed that in the V346 Nor system, there is an
inner Keplerian disk surrounded by a infalling-rotating pseudo-disk,
and an extended envelope, including a relatively narrow outflow
cavity. These components are typically found in Class 0/I protostellar
systems as well. At the beginning, the kinematics are dominated by the
infalling motion within the envelope, but then the envelope starts a
combination of rotation and infall with conserved angular momentum, as
demonstrated by measurements of the $v_{\rm rot} \sim r^{-1}$ velocity
profile (e.g., VLA~1623~A, \citealt{murillo2013}). In more evolved
cases, a disk in Keplerian rotation is present as well (e.g.,
L1489~IRS, \citealt{hogerheijde2001,yen2014}).

Table~\ref{tab:class} shows the envelope and disk parameters of some
well-studied Class 0/I protostars from the literature. A comparison
with V346~Nor shows that the main characteristics of our target fall
within the parameter range defined by these protostars. Concerning the
velocity field in V346~Nor, a Keplerian disk is already present, but
there are still signs of infall in the outer parts as indicated by the
sub-keplerian rotation. This suggests that among protostars, our
target represents an intermediate evolutionary state. In summary, the
measured kinematic properties of V346~Nor enabled us to determine its
evolutionary status, which could not be directly derived from its SED
and bolometric temperature and luminosity due to the outburst.

\begin{deluxetable*}{cccccc}
\tablecaption{Stellar, disk and envelope properties of Class 0/I
  sources and V346~Nor.\label{tab:class}} \tablehead{ &
  \colhead{L1527~IRS} & \colhead{VLA~1623~A} & \colhead{HH~212} &
  \colhead{L1551~IRS~5} & \colhead{V346~Nor} }
\startdata
$M_*$ ($M_{\odot}$)	  & 0.19	& 0.22       & 0.18    & 0.1	     & 0.1	\\
$R_{\rm disk}$ (au)	  & 150 	& 180	     & 120     & 160	     & 350	 \\
$M_{\rm env}$ ($M_{\odot}$) & 0.9 	 & 0.8        & 0.1	& 0.06         & 0.23   \\
$R_{\rm env}$ (au)	 & 10\,000     &	     & 2000    & 1200	    & 6300	\\
$\dot{M}_{\rm infall}$ ($M_{\odot}$\,yr$^{-1}$) & 7$\times$10$^{-6}$ & & 5$\times$10$^{-6}$ & 6.4$\times$10$^{-6}$ & 6$\times$10$^{-6}$ \\
Reference	       & (1)	      & (2)	   & (3)       & (4)	  & (5)     \\
\enddata
\tablerefs{(1) \citet{tobin2012}; (2)
  \citet{murillo2013}; (3) \citet{lee2014}; (4) \citet{momose1998};
  (5) this work.}
\end{deluxetable*}

\subsection{Comparison to other FUors}

There are several high spatial resolution millimeter wavelength
observations in the literature for FUors. In the following, we
summarize their main findings and compare V346~Nor with them.

Z~CMa is a complicated binary system of a FUor of 3$\,M_{\odot}$ and a
Herbig Be star of 16$\,M_{\odot}$, with a separation of 105\,au
\citep{koresko1991, vandenancker2004}. \citet{alonso-albi2009} found a
massive and compact envelope in the form of a toroid extending from
2000\,au to 5000\,au. To reproduce the SED, they needed a disk with an
outer radius of 180\,au, and they also mention the possibility of an
outer, cold, spherical envelope. According to our APEX $^{12}$CO(3--2)
data (K\'osp\'al et al., in prep.), the CO emission has a deconvolved
size of about 30\,000\,au, corroborating the claim of
\citet{alonso-albi2009}. Z~CMa is known to drive a molecular outflow
and has associated jets and HH objects \citep{poetzel1989, evans1994}.

L1551~IRS~5 is a well-studied prototypical embedded protostar which
underwent an FU~Orionis-type outburst, and which is also a binary
object with a $\sim$50\,au separation, including circumstellar disks,
a circumbinary structure, and a large-scale envelope as well
\citep{looney1997}. L1551~IRS~5 drives a strong molecular outflow
discovered by \citet{snell1980}. \citet{fridlund2002} give a summary
of the millimeter molecular line observations of the object, and
present new observations indicating a large ($\sim$7000\,au radius)
rotationally supported flattened structure, embedded in an even more
extended ($\sim$10\,000\,au radius) envelope. There are signs for
infall of gas towards this disk-like structure \citep[see
  also]{ohashi1996, momose1998}. \citet{chou2014} suggests a Keplerian
disk of radius 64\,au and mass 0.07$\,M_{\odot}$ around the
protobinary with a central mass of 0.5$\,M_{\odot}$, and a flattened,
rotating-infalling envelope of 1$\,M_{\odot}$ that connects to the
inner Keplerian disk.

With a stellar mass of $\sim$0.25\,$M_{\odot}$, V2775~Ori is one of
the lowest luminosity and lowest mass FUors
\citep{carattiogaratti2011}. Using ALMA, \citet{zurlo2017} found a
very low-mass disk around the object ($>$3$\,M_{\rm Jup}$).  They also
detected an outflow cavity marked by the CO emission of the material
in the cavity walls, with an opening angle of about 30$^{\circ}$. The
half-axis length of the outflow cavity is about 2500\,au, so there
should be an envelope of at least this radius around the system.

HBC~722 is a young eruptive star which has been in outburst since
2010. In quiescence, it was a normal T~Tauri star with a K7-type
(0.5--0.6\,$M_{\odot}$) central star and a disk which remains
undetected at millimeter wavelength, giving an upper limit of
0.01--0.02\,$M_{\odot}$ for the total disk mass
\citep{miller2011,kospal2011a,dunham2012b}. \citet{kospal2016}
discovered a flattened molecular gas structure with a diameter of
1700~au and mass of 0.3\,$M_{\odot}$ centered on HBC~722. No outflows
are driven by the FUor (see also \citealt{dunham2012b}).

HBC\,494 is an embedded object showing FUor characteristics associated
with the Reipurth\,50\,N nebula
\citep{reipurth1986,reipurth1997b}. Based on ALMA observations,
\citet{ruiz-rodriguez2017} found a wide-angle outflow extending to
8000\,au. The opening angle of the outflow cavity is
$\sim$150$^{\circ}$, the widest known cavity among Class\,I
protostars. At low velocities, the material at the base of the cavity
shows infall toward the central object.

These previous sources are mostly young, embedded
objects. Interestingly, the prototype object, FU~Ori is different from
them. It is a relatively wide separation binary (225\,au,
\citealt{wang2004}).  ALMA observations by \citet{hales2015} suggest
disks around both components with sizes of $<\,$45\,au. The small disk
sizes were confirmed by \citet{liu2017}. They detected no extended
emission on spatial scales smaller than 3000\,au, however, in our APEX
data we resolved the $^{12}$CO(3--2) emission, and found an envelope
with a radius of about 8000\,au around the system (K\'osp\'al et al.,
in prep.). FU~Ori is not alone among FUors with having a more tenuous
envelope. Similarly small-mass gas envelopes were found around Bran~76
and HBC~687 \citep{kospal2017b}.

The wide variety of envelope properties observed in FUors demonstrates
that the eruptive phenomenon is present during the whole evolution
from the earliest protostellar phases until the disk-only state. This
was also the conclusion of the APEX CO survey of \citet{kospal2017b},
where we found that the observed FUor sample could be divided into
high ($>$0.3$\,M_{\odot}$) and low (0.01--0.02$\,M_{\odot}$) envelope
mass classes. We suggested that this division is also consistent with
the findings of \citet{quanz2007}, who proposed that FUors evolve from
an embedded state showing 10$\,\mu$m silicate absorption to a more
evolved state showing 10$\,\mu$m silicate emission. The increasing
visibility of the inner part of the system, reflected in the silicate
feature, is driven by the subsequent eruptions associated with strong
outflows. \citet{green2006} argued that the opening angle of the
outflow cavity would increase as the FUor evolves.

Our ALMA data show that V346~Nor is surrounded by an envelope with a
total gas mass of 0.23$\,M_{\odot}$. This result indicates that it
belongs to the embedded subgroup of FUors. We measured opening angles
of the outflow cavity between 40 and 80$^{\circ}$, which is relatively
wide compared to some Class\,0/I protostars but narrower than that of
HBC~494. These results suggest that V346~Nor is still at a relatively
early evolutionary phase, although its envelope properties and
kinematics suggest that the outbursts already had an impact on the
circumstellar material and it is on its way toward the T Tauri phase.

\subsection{Infall/accretion mismatch}

\citet{momose1998} studied the gas kinematics around L1551~IRS~5, and
found that the outer parts of the envelope show infall with conserved
angular momentum, which shifts to a rotationally supported Keplerian
disk in the inner regions. They explain that in such case, the infall
rate can be estimated from the mass of the envelope and the infall
velocity as follows:
\begin{equation}
  \dot{M}_{\rm infall} = \frac{3}{2} \left( \frac{M_{\rm env}}{R_{\rm env}} \right)
  \left( \frac{2GM_*}{R_{\rm env}} \right)^{1/2}
\end{equation}
Using the similarity of the velocity field observed in V346~Nor to
that of L1551~IRS~5, we estimated the infall rate from the envelope
onto the disk in V346~Nor using the equation above, obtaining
6$\times$10$^{-6}\,M_{\odot}$\,yr$^{-1}$. Taking into account the
uncertainty in the stellar mass, envelope mass, and envelope size, we
estimate that this number is correct within a factor of 3. In the
following, we relate this value to the quiescent and outburst
accretion rates from the disk onto the star.

Using a simple accretion disk model, \citet{kospal2017a} estimated the
accretion rate at multiple epochs throughout the outburst. While this
gives fairly good indication on how the accretion rate changes with
time in V346~Nor, its absolute value is more uncertain, as it depends
strongly on the inclination (the model assumes a geometrically thin
disk, which is probably not true in such a young system). In order to
have independent estimates for the accretion rate in V346~Nor, we can
calculate the bolometric luminosity by integrating the area under the
outburst SED at peak brightness plotted in $\nu{}F_{\nu}$, assume that
it is dominated by the accretion luminosity (good assumption for a
FUor in outburst), and calculate $\dot{M}$ from the following
expression, which assumes that all gravitational potential energy is
radiated away \citep{evans2009}:
\begin{equation}
  \dot{M} = \frac{L_{\rm bol}R_*}{GM_*}
\end{equation}
The SED gives an $L_{\rm bol}$ = 160\,$L_{\odot}$, which gives an
accretion rate of $\dot{M} = 1.5\times10^{-4}\,M_{\odot}$\,yr$^{-1}$
using $M_*=0.1\,M_{\odot}$ and $R_* = 3\,R_{\odot}$. This is close to
the value given by \citet{kospal2017a} for the maximal accretion rate,
which was $\dot{M} = 10^{-4}\,M_{\odot}$\,yr$^{-1}$. Another way to
estimate the accretion rate is to use the Br$\gamma$ line luminosity
as an accretion rate tracer. \citet{connelley2010} measured a
Br$\gamma$ equivalent width of 2.26\,$\AA$. Using the K$_S$-band
photometry of 7.21\,mag to calibrate the line flux, we obtained
1.3$\times$10$^{-16}$\,W\,m$^{-2}$ for the Br$\gamma$ line flux (or
0.002\,$L_{\odot}$ for the line luminosity). Using the empirical
relationship between the Br$\gamma$ line luminosity and the accretion
luminosity from \citet{mendigutia2011}, we obtained $\dot{M} =
9\times10^{-5}\,M_{\odot}$\,yr$^{-1}$, by correcting for a reddening
of $A_V$ = 14\,mag \citet{kospal2017a}. This may be lower than what we
obtained from the SED because at the time of the Br$\gamma$
measurement of \citet{connelley2010}, V346\,Nor might have already
started its fading due to a temporary decrease in the accretion.

\citet{kospal2017a} claimed that the accretion rate during the 2010
minimum of V346~Nor was at least a factor of 100 less than during its
high accretion phase. This means it should be less than about $\dot{M}
= 10^{-7}-10^{-6}\,M_{\odot}$\,yr$^{-1}$. There is no other way to
estimate the accretion rate pre-outburst, because the full optical-mm
SED is not available before 1980, neither is there a Br$\gamma$
measurement from that time. Comparing this with the envelope infall
rate we calculated, 6$\times$10$^{-6}\,M_{\odot}$\,yr$^{-1}$, there is
a hint for a mismatch between the two values. The infall rate is
larger, suggesting that the disk receives more material from the
envelope than it can transfer onto the star with the quiescent
accretion. It was speculated in the literature that the piling up of
material in such a situation is the explanation for the FUor
phenomenon \citep{bell1994}. In V346~Nor, we might witness the first
observational support that the mismatch between the infall and
accretion rates indeed leads to FUor outbursts.

The large infall rate of 6$\times$10$^{-6}\,M_{\odot}$yr\,$^{-1}$, has
another interesting implication. According to the numerical
calculations of \citet{ohtani2014}, bursts induced by
magneto-rotational instability (MRI) only happen below a critical
value of 3$\times$10$^{-6}\,M_{\odot}$\,yr$^{-1}$, above which bursts
are suppressed and accretion proceeds in a quasi-stable manner. The
critical value may be higher, as suggested by simulations of the
MRI-induced bursts by \citet{bae2014} and
\citet{zhu2008}. Nevertheless, there always exists a threshold value
for the mass infall rate above which the MRI-induced bursts
cease. Such threshold does not exist in the disk fragmentation model
of \citet{vorobyov2010}, in which higher infall rates only promote
more vigorous disk fragmentation. In the following, we discuss whether
this model can explain the outburst history of V346~Nor.

In the disk fragmentation scenario, luminosity outbursts are caused by
gaseous fragments forming via disk gravitational fragmentation and
in-spiralling onto the star due to the loss of angular momentum caused
by the gravitational interaction with spiral arms
\citep{vorobyov2015}. Depending on the ability of the fragments to
withstand the tidal torques when approaching the star, two types of
bursts can occur. In isolated bursts, the fragment is accreted almost
intact, producing a well-defined burst. In clustered bursts, the
fragment is stretched into a knotty filament when approaching the
star, thus producing a series of closely packed bursts of varying
amplitude separated from each other by several decades. The light
curve of V346~Nor indicated a temporary halt in the accretion around
2010, followed by a re-brightening, all happening within a few years
\citep{kraus2016, kospal2017a}. This time span is too short to allow
for a disk refill and transport of matter from the outer disk (scales
of 100\,au). The character of the V346~Nor outburst is reminiscent of
the clustered burst phenomenon, although the mass of the disk we found
in our ALMA observations (0.01--0.07$\,M_{\odot}$) may be too low for
gravitational fragmentation. We estimated the Toomre Q parameter for
our disk following Equation (2) of \citet{kratter2016}, and found that
the disk can either be graviationally unstable or stable depending on
the disk mass we use in the 0.01--0.07$\,M_{\odot}$ range. Other
physical possibilities for the interrupted outburst of FU~Ori-type
stars are reviewed in \citet{ninan2015}. Finally, we note that in the
gravitation fragmentation scenario, the luminosity between the closely
packed bursts rarely falls below the pre-burst value (Fig.~12 in
\citealt{vorobyov2015}), suggesting that the pre-outburst mass
accretion rate in V346~Nor was even lower than that during the 2010
minimum.


\section{Summary}

We present the first ALMA observations of the FUor-type young
outbursting star V346~Nor at 1.322\,mm continuum and in the J=2--1
lines of $^{12}$CO, $^{13}$CO, C$^{18}$O. In order to image the
emission correctly at all spatial scales, we combined interferometric
data from the 12\,m and 7\,m arrays with Total Power observations. Our
main results and conclusions are the following:

\begin{itemize}
\item We detected a fairly compact continuum source coinciding with
  the near-infrared stellar location, surrounded by more extended,
  fainter emission out to 6300\,au. The dust mass in the compact
  source within 210\,au is 7$\times$10$^{-4}\,M_{\odot}$.
\item We detected extended CO emission in all three isotopologues. The
  spectra show profiles similar to what is typical for infalling
  envelopes around highly embedded objects.
\item Within the central beam, the $^{13}$CO and C$^{18}$O spectra
  exhibit double-peaked profiles, symmetric around the systemic
  velocity of $-$3.55\,km\,s$^{-1}$, suggesting rotation.
\item A spectro-astrometric analysis of the C$^{18}$O data cube
  revealed that from 350 to 700\,au, the radial velocity profile is
  consistent with a pseudo-disk, while the inner 350\,au more
  resembles a Keplerian disk around a 0.1$\,M_{\odot}$ central
  star. The total gas mass within 350 and 700\,au is 0.01 and
  0.03$\,M_{\odot}$, respectively.
\item The channel maps at large velocities reveal structures that we
  interpret as the walls of an outflow cavity, and measure a
  relatively wide opening angle of 40--80$^{\circ}$.
\item A comparison of V346~Nor with well-studied Class 0/I protostars
  show that the main characteristics of our target fall within the
  parameter range defined by these protostars.
\item Comparison with FUors having similar high spatial resolution
  millimeter CO data indicates that it belongs to the embedded
  subgroup of FUors.
\item We determined an infall rate from the envelope onto the disk of
  6$\times$10$^{-6}\,M_{\odot}$\,yr$^{-1}$, which is a factor of few
  higher than the quiescent accretion rate from the disk onto the
  star, hinting for a mismatch between the infall and accretion
  rates. This is the first observational support for such mismatch in
  a FUor, previously invoked to explain FUor outbursts.
\item The computed infall rate cannot exclude the MRI-driven outburst
  mechanism. Accretion of clumps in a gravitationally fragmenting disk
  is also a possibility, although the measured disk mass may be too
  low for this process.
\end{itemize}


\acknowledgments

The authors acknowledge support by Allegro, the European ALMA Regional
Center node in The Netherlands, and expert advice from Yanett
Contreras in particular. This work was supported by the Momentum grant
of the MTA CSFK Lend\"ulet Disk Research Group. E.~I.~Vorobyov
acknowledges support from the Austrian Science Fund (FWF) under
research grant I2549-N27. This paper makes use of the following ALMA
data: ADS/JAO.ALMA\#2013.1.00870.S. ALMA is a partnership of ESO
(representing its member states), NSF (USA) and NINS (Japan), together
with NRC (Canada) and NSC and ASIAA (Taiwan) and KASI (Republic of
Korea), in cooperation with the Republic of Chile. The Joint ALMA
Observatory is operated by ESO, AUI/NRAO and NAOJ.

\facility{ALMA}.


\appendix

\section{Channel maps of V346~Nor}

\begin{figure*}
\includegraphics[angle=0,scale=0.9]{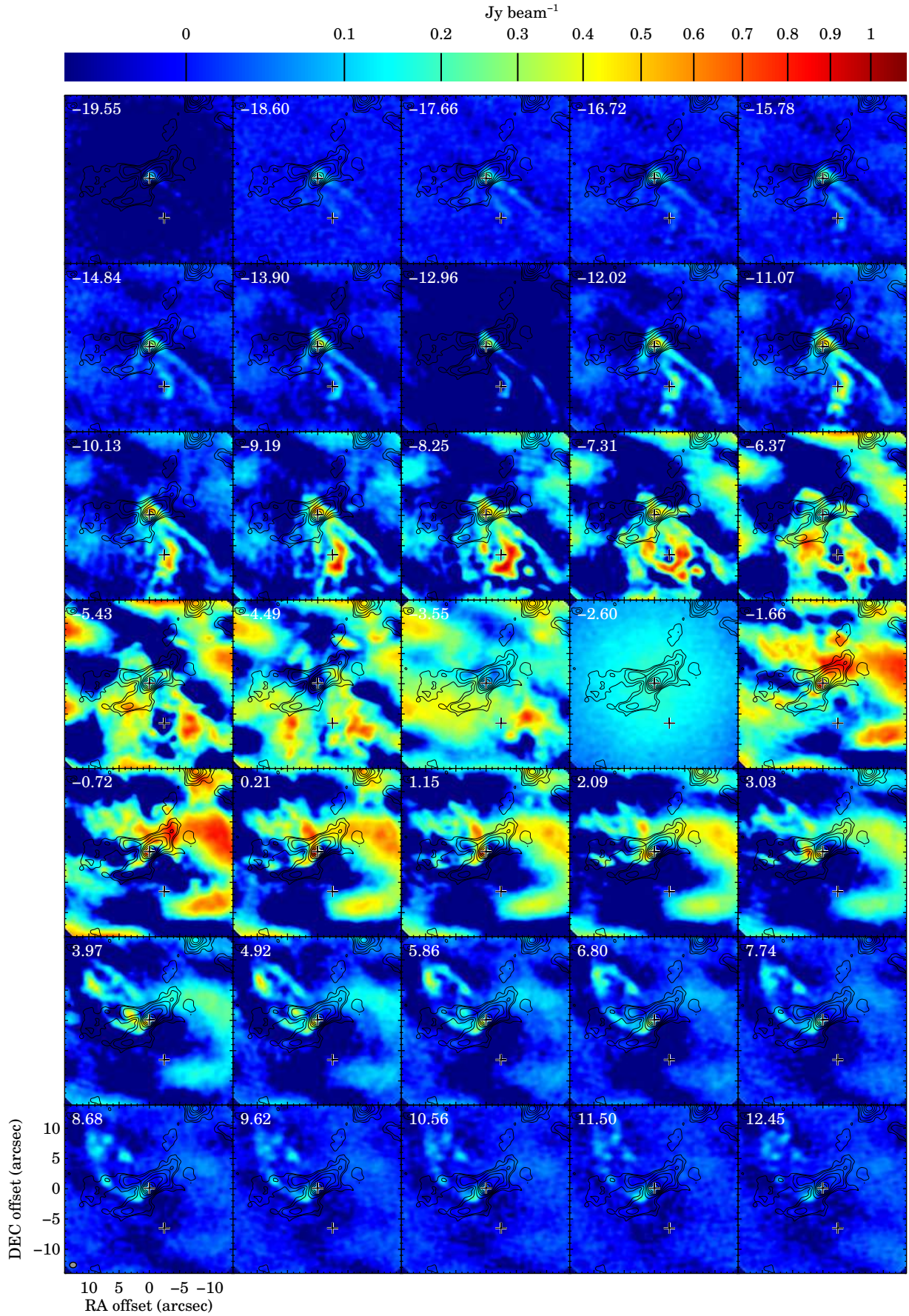}
\caption{ALMA $^{12}$CO channel maps of of V346~Nor. The beam is
  indicated in the lower left panel. The numbers in the upper left
  corner of each panel display the velocity in km\,s$^{-1}$.
  \label{fig:channelmaps_12co}}
\end{figure*}

\begin{figure*}
\includegraphics[angle=0,scale=0.9]{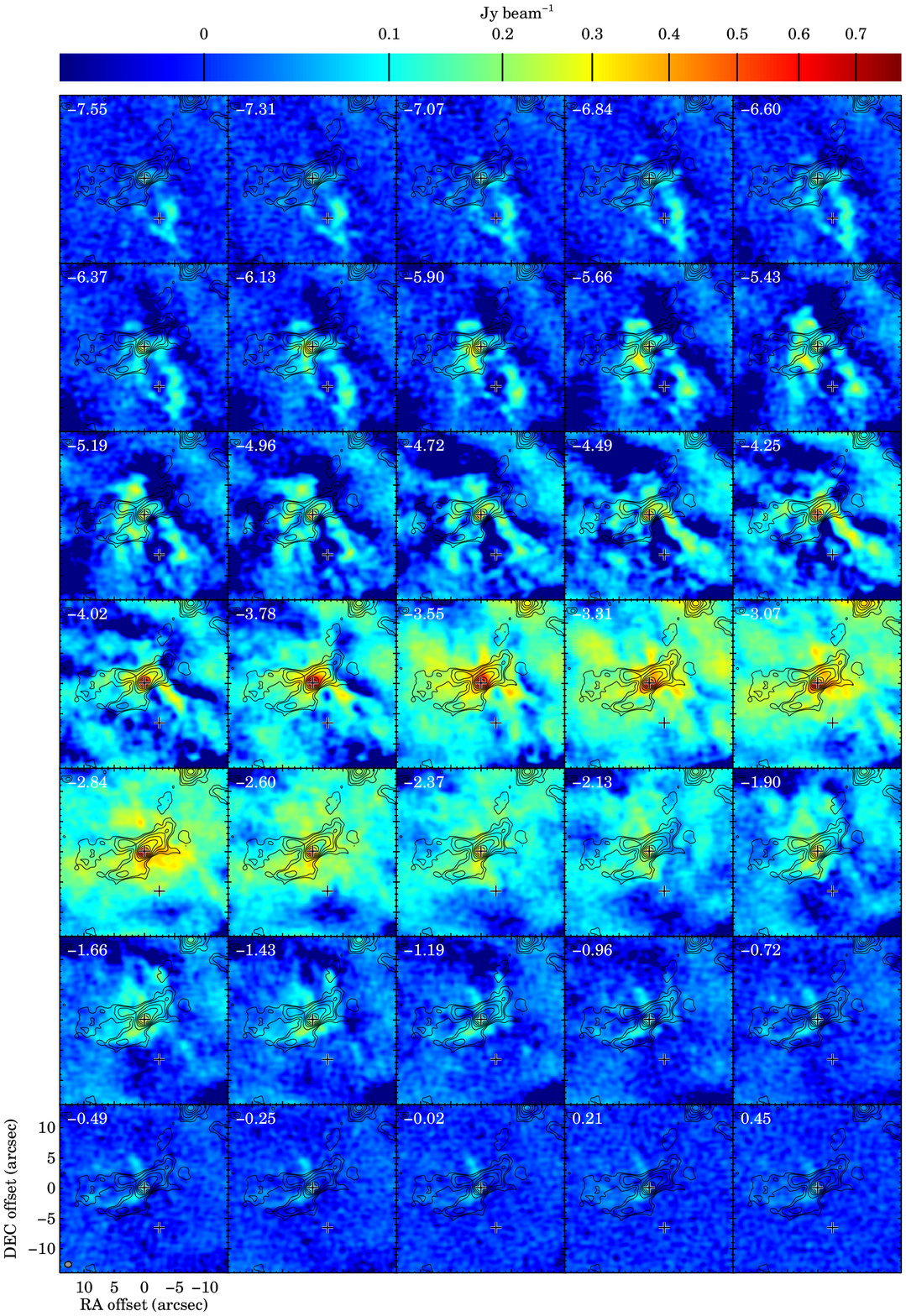}
\caption{ALMA $^{13}$CO channel maps of of V346~Nor. The beam is
  indicated in the lower left panel. The numbers in the upper left
  corner of each panel display the velocity in km\,s$^{-1}$.
  \label{fig:channelmaps_13co}}
\end{figure*}

\begin{figure*}
\includegraphics[angle=0,scale=0.9]{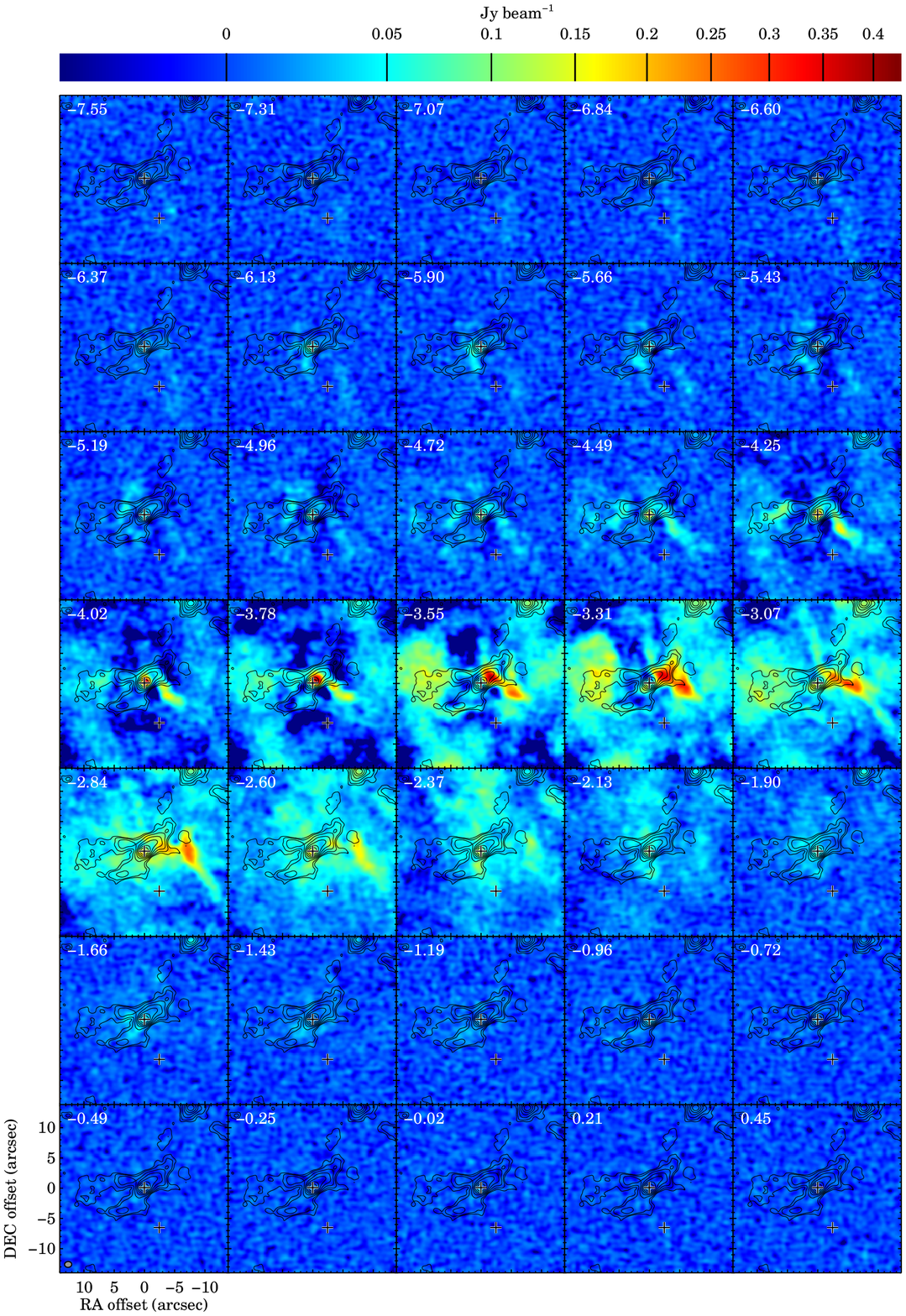}
\caption{ALMA C$^{18}$O channel maps of of V346~Nor. The beam is
  indicated in the lower left panel. The numbers in the upper left
  corner of each panel display the velocity in km\,s$^{-1}$.
  \label{fig:channelmaps_c18o}}
\end{figure*}


\bibliography{paper}{}


\end{document}